# Polarimetry of Water Ice Particles Providing Insights on Grain Size and Degree of Sintering on Icy Planetary Surfaces

**O. Poch[1]†\*, R. Cerubini[1], A. Pommerol[1], B. Jost[1], N. Thomas[1]**

[1]NCCR PlanetS, Physikalisches Institut, Universität Bern, Bern, Switzerland

Corresponding author: Olivier Poch (olivier.poch@univ-grenoble-alpes.fr)

†Current address: Univ. Grenoble Alpes, CNRS, CNES, IPAG, 38000 Grenoble, France

**Key Points:**

- Phase curves of the scattered light and degree of linear polarization were measured on surfaces of well-characterized water ice particles

- Influences of ice particles size and degree of metamorphism on the phase curves are described. Frost shows resonances (Mie oscillations)

- The laboratory data suggest average icy grains diameters possibly ~40-400 μm on Europa but only few micrometers on Enceladus and Rhea

## Abstract

The polarimetry of the light scattered by planetary surfaces is a powerful tool to provide constraints on their microstructure. To improve the interpretation of polarimetric data from icy surfaces, we have developed the POLarimeter for ICE Samples (POLICES) complementing the measurement facilities of the Ice Laboratory at the University of Bern. The new setup uses a high precision Stokes polarimeter to measure the degree of polarization in the visible light scattered by surfaces at moderate phase angles (from 1.5 to 30°). We present the photometric and polarimetric phase curves measured on various surfaces made of pure water ice particles having well-controlled size and shape (spherical, crushed, frost). The results show how the amplitude and the shape of the negative polarization branch change with the particles sizes and the degree of metamorphism of the ice. We found that fresh frost formed by water condensation on cold surfaces has a phase curve characterized by resonances (Mie oscillations) indicating that frost embryos are transparent micrometer-sized particles with a narrow size distribution and spherical shape. Comparisons of these measurements with polarimetric observations of the icy satellites of the Solar System suggest that Europa is possibly covered by relatively coarser (~40-400 μm) and more sintered grains than Enceladus and Rhea, more likely covered by frost-like particles of few micrometers in average. The great sensitivity of polarization to grain size and degree of sintering makes it an ideal tool to detect hints of ongoing processes on icy planetary surfaces, such as cryovolcanism.

## Plain Language Summary

Light can be considered as an oscillating wave that propagates in space. When the light from the Sun is reflected by a planetary surface, the direction of its oscillation partially changes. The information about the direction of the oscillation is known as the "polarization" of the light. In this study, we show how the polarization of the visible light reflected by surfaces made of water ice grains changes with the size of these grains and the freshness of the ice surfaces. By comparing our results to astronomical observations, we find that the surface of Jupiter's moon Europa is made of much larger ice grains (table salt size-like) than the surfaces of Saturn's moons Enceladus and Rhea (finer than icing sugar). This striking difference is probably due to different environmental conditions transforming these surfaces.

## 1 Introduction

The unpolarized star light gets partially polarized when scattered by a planetary surface. In addition to the properties of the surface, the degree of polarization of the scattered light depends on the solar phase angle α (angle between the Sun, the object and the observer) and the wavelength at which the surface is observed. The curve of the degree of linear polarization versus phase angle, called polarimetric phase curve, is very sensitive to the size, shape, structure and composition (complex refractive index) of the scattering particles. As a consequence, observing the polarization of the scattered light can significantly complement observations performed in total light intensity, providing additional constraints to interpret remote sensing observations of Earth, Solar System bodies or extra-solar objects.

At relatively small phase angles, the scattered light exhibits a surge of intensity called the opposition effect and is partially linearly polarized parallel to the scattering plane (the plane containing the Sun, the object and the observer), a feature called the "negative polarization branch" (from 0 to about 30°). The term "negative" comes from the definition of the polarization of the scattered light $P$ (usually expressed in percentage) as

$$P = \frac{Q}{I} = \frac{I_\perp - I_\parallel}{I_\perp + I_\parallel}$$

with $Q$ the second Stokes parameter, $I$ the total intensity of light, and $I_\perp$ and $I_\parallel$ the components of light intensity in the direction orthogonal and parallel to the plane of scattering respectively. Generally, the negative polarization branch (NBP) displays a parabolic shape between 0 and 30° with a minimum of polarization ($P_{min}$) at a small phase angle $\alpha_{min}$ and then changes of sign at an angle $\alpha_{inv}$, typically around 20°, to become positive at larger phase angles. For some high-albedo objects, this negative branch has been found to be bimodal, with a second minimum below 2° phase angle, a phenomenon called the "polarization opposition effect" (POE) (Rosenbush et al., 2015 and references herein).

The origin of these ubiquitous features is not yet fully understood theoretically, and a numerical inversion method to interpret both the photometric and polarimetric phase curves observed from planetary surfaces is not yet available (Muinonen et al., 2015). Therefore, laboratory measurements of phase curves of analogues of planetary surfaces, made of materials and textures similar to (extra-)terrestrial ones, provide essential references to interpret observations and validate theoretical models currently in development.

Since the early twentieth century, polarimetric phase curves measured in the laboratory have been used to infer the surface properties of airless bodies of the Solar System, revealing, for examples, the high porosity of the lunar regolith (Lyot, 1929, 1934) or the presence of fine iron oxide dust at the surface of Mars (Dollfus, 1957). Ground-based polarimetric observations of Solar System bodies collected over many years (Kolokolova et al., 2015), constitute a precious data set having the potential to reveal the micro-texture of these distant planetary surfaces. A large number of measurements of analogues of planetary regolith were compiled by (Geake & Dollfus, 1986; Nelson et al., 2018; Ovcharenko et al., 2006; Shkuratov, 2002; Shkuratov et al., 2006) among others, and compared with observations. Polarimetry is also used in Earth remote sensing to characterize mainly the atmospheric aerosols or the surfaces. Even for the retrieval of aerosols properties, a good knowledge of the polarization from the surface is needed because it contributes to the signal and has to be removed (Gatebe et al., 2010; Herman et al., 1997). Various laboratory and field measurements campaigns were carried out to measure the polarization of typical terrestrial surfaces (Peltoniemi et al., 2009, 2015). Recently, a database of the polarimetric properties of various terrestrial surfaces, generated from the data of the POLDER instrument onboard the PARASOL satellite, has been released (Breon & Maignan, 2017).

In the present study, we aim at improving the knowledge of the polarimetry of surfaces composed of water ice. Ice particles are widespread on the moons of outer planets, as well as on terrestrial planets, comets, Trans-Neptunian Objects (TNO) and even on some asteroids, whose polarimetric phase curves need to be interpreted (Rosenbush et al., 2015). Moreover, the characterization via remote sensing of the micro-structure and melting state of surfaces covered by snow on Earth is required to constrain their influence on surface albedo and climate (Kelly & Hall, 2008).

Despite this context, the polarization of light scattered by snow and frost has only been the object of a limited number of studies. It was first measured by Lyot (1929) and Dollfus (1957), who observed relatively low values of polarization and variations of the phase curves with the degree of melting. Follow-up measurements done in the context of astronomy are mainly limited to small phase angles (between 0 to 30°) and visible wavelengths, with high polarimetric accuracy (typically 0.05%). Steigmann (1993) measured frost covered by varying amount of fine silicate dust in an effort to match the polarization phase curve of Callisto. Dougherty and Geake (1994) revealed that sub-micrometer-sized water ice frost particles exhibit much deeper NPBs than coarser ice grains. Shkuratov and Ovcharenko (2002) found no deep negative branch of polarization between 0 and 3° phase angle for fresh fallen snow made of 50×500 µm elongated particles. Measurements performed in the context of Earth remote sensing cover much larger phase angles and wavelengths (up to 2.5 µm), but with lower polarimetric accuracy (typically 1%). Various field measurements characterized the polarization phase curves and spectra of natural snow of different textures (new snow, melting snow, melt freeze crust, surface hoar etc.) and found dependences with grain sizes and shape (Leroux et al., 1999; Lv & Sun, 2014; Peltoniemi et al., 2009; Sun & Zhao, 2011; Tanikawa et al., 2014).

However, no systematic study has been done on ice particles of well-controlled shape, size and composition. At the University of Bern, we have developed a series of Setups for the Preparation of Icy Planetary Analogues (SPIPA) that are used to prepare spherical water ice particles, pure or mixed with contaminants, in a reproducible way. In previous works, we have measured their spectro-photometric properties: phase curves and bidirectional reflectance distribution functions (BRDF) using the PHIRE-2 instrument (Jost et al., 2016; Yoldi et al., 2015), and reflectance spectra using the SCITEAS setup (Poch et al., 2016). In the present study, we describe the first results of a new polarimetric setup developed in our group, called POLICES for POLarimeter for ICE Samples.

The manuscript is organized as followed. In section 2.1, the POLICES setup and its calibration are described. In section 2.2, we explain the preparation of the ice samples, made of particles of different shapes and sizes. Section 3 presents the measured photometric and polarimetric phase curves of the samples and their temporal evolution as the ice undergoes metamorphism. In sections 4.1, 4.2 and 4.3 these results are interpreted using numerical modeling and compared to previous works. In section 4.4, the laboratory measurements are directly compared to polarimetric observations of the icy satellites Europa, Enceladus and Rhea, to infer the size of the ice particles covering their surfaces. Finally, section 5 summarizes the main results and findings. As supplementary information documents, additional figures (noted as S1, S2, S3 and S4) are provided, as well as ASCII text files containing all the data presented, which can be used as references to test theoretical models and predict or interpret polarimetric observations.

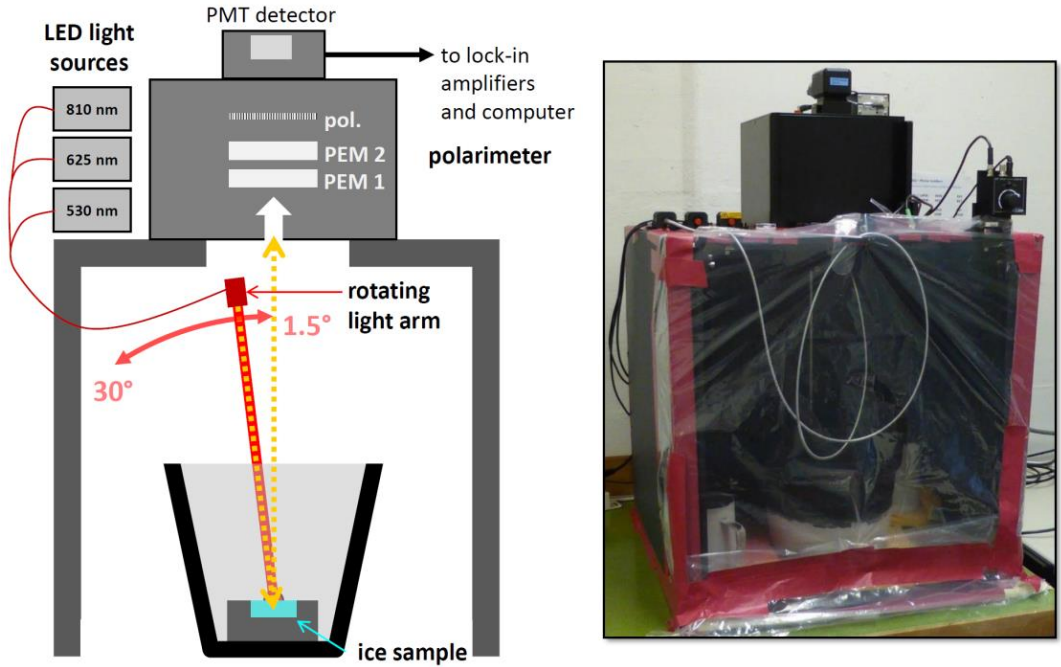

**Figure 1.** The POLICES setup, consisting of a hermetical enclosure allowing the measurements of photometric and polarimetric phase curves of icy samples under dry and cold conditions. A motorized "light arm" illuminates the sample surface with monochromatic light at incidence angles from 1.5 to 30°, and a polarimeter analyses the scattered light at a fixed emergence of 0°.

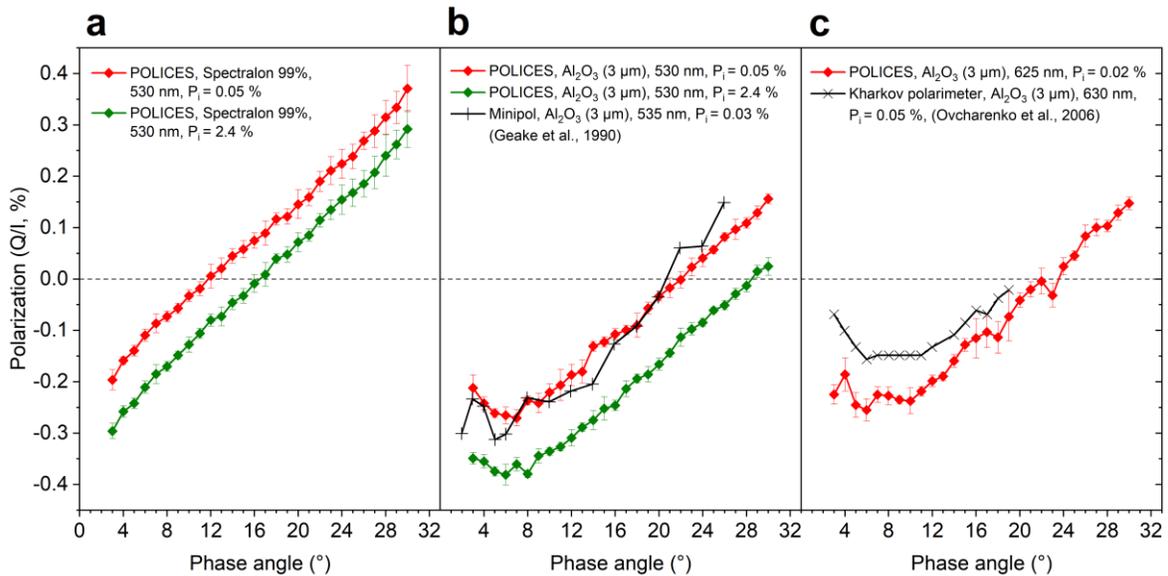

**Figure 2.** Measurements of **(a)** Spectralon with and without depolarizer, **(b)** surface made of 3µm-diameter aluminum oxide (WCA3) with and without depolarizer, and as measured by Geake & Geake (1990), and **(c)** same surface measured with depolarizer at 625 nm compared to measurement by Ovcharenko et al. (2006) at 630 nm. The red phase curves are available in supporting information Data Set S1.

## 2 Materials and Methods

### 2.1 The POLarimeter for ICE Samples (POLICES)

#### 2.1.1    Sample environment and data acquisition

The POLICES setup is a gonio-polarimeter, enabling the measurement of the intensity and polarization state of the light yscattered by icy surfaces (Figure 1). One important issue when aiming to measure well-controlled icy surfaces is to avoid the condensation of water vapour forming a layer of frost on these surfaces. Therefore, the POLICES setup consists of a cubic enclosure 60×60×60 cm which can be sealed and purged with nitrogen in order to reduce the relative humidity and prevent the formation of water cloud and frost on the sample. Once prepared, the sample of ice particles at about 173 K is put in an insulative foam box (Magic Touch$^{TM}$ 2) 20×20 cm large and 15 cm high, pre-cooled with liquid nitrogen and consequently filled with dry nitrogen gas. The cold box containing the sample, closed with a lid, is placed in the POLICES enclosure together with several containers filled with about 5 liters of boiling liquid nitrogen, ensuring the purge and cooling of the enclosure which is then sealed for the whole duration of the measurements. A sensor (Thorlabs TSP01), positioned at the height of the cold box, measures the relative humidity in the enclosure. When it reaches a minimum (typically about 25%), the lid of the box is removed thanks to a glove box interface. The measurement starts immediately after the removal of the lid. Under these conditions, the temperature of the sample, measured with a Pt100 sensor, slowly increase from about 173 K to 220 K during one hour. By monitoring of the sample surface using a CCD camera, we can confirm that no water cloud nor frost were formed on the sample during the measurements.

The sample surface is positioned in the horizontal plane containing the rotation axis of a motorized arm holding a collimated light source. The sample can thus be illuminated at incidence angles ranging from 1.5° to 30°, with a spot diameter of about 15 mm on its surface. Monochromatic light produced by one of three power-controlled fiber-coupled LEDs at 530, 625 or 810 nm (Thorlabs M530F2, M625F2 or M810F2) can be alternately sent to the collimator (Thorlabs RC08SMA-P01). In the current study, mostly measurements at 530 nm are presented. The polarization (Q/I) of the incident light was measured by orienting the collimator in the optical axis of the polarimeter and it was of 2.4 % at 530 nm. This degree of polarization is high compared to that of the Sun light illuminating icy surfaces of airless Solar System bodies, of the order of 0.01 % or less (Clarke & Fullerton, 1996; Kemp et al., 1987). The use of a Liquid Crystal Polymer Depolarizer (Thorlabs DPP25-A) positioned in front of the collimator output reduces this polarization to 0.05 %. However, during the measurements presented here, the ice samples were illuminated without depolarizer so we had to investigate how an incident polarization of 2.4 % influence the polarization of their scattered light, as explained below in section 2.

The light scattered by the surface sample is analyzed by a precision Stokes polarimeter (Dual PEM II/FS42-47, Hinds Instruments) placed on top of the enclosure and whose optical axis is fixed at 0° emergence, in the normal direction of the sample surface. In this configuration, the angle between the incidence and emission directions is the "phase angle". The vertical plane containing the incidence and emergence directions is the "scattering plane" or the "principal plane". The field of view of the polarimeter covers a 122 mm diameter disk on the sample surface, including the 10 mm light spot. The angular resolution of the system is of ±1° phase angle.

The light entering the polarimeter through a 23 mm aperture encounters a first photoelastic modulator (PEM) oriented parallel to the principal plane and modulated at resonance frequency $f_1$ = 42 kHz, followed by a second PEM oriented 45° to the first one and modulated at $f_2$ = 47 kHz. After the two PEMs, the light goes through a polarizer with passing axis oriented at 22.5° to each modulator, and finally a lens and a high sensitivity photomultiplier detector (PMT Hamamatsu R928). By setting the amplitude of the PEMs so that they induce a retardation of ±λ/2 for the component of the electric field parallel to their vibration axis, the system of PEMs and polarizer converts the information on the orientation of the electric field into a modulation of the light intensity reaching the detector. This detector generates an electric signal with a RMS voltage (DC) proportional to the light intensity (for a given gain) and an AC component modulated by the PEMs. Three lock-in amplifiers are used to demodulate the AC signal at frequencies equal to $2f_1$, $2f_2$ and $1f_1$. The reduced Stokes parameters Q/I and U/I (for linear polarization) are obtained respectively by the $2f_1$ and $2f_2$ modulation frequencies of the PEMs, while the reduced Stokes parameter V/I (for circular polarization) is measured by the $1f_1$ modulation frequency of the first PEM. This system allows the light intensity and the three reduced Stokes parameters Q/I, U/I and V/I to be measured simultaneously in one millisecond. In practice, the result reported by the lock-in amplifiers is an average over 60 ms. The resolution at which the lock-in amplifiers demodulate the signal is of $10^{-4}$. For this study, we only discuss the values of Q/I to compare with astronomical observations (Equation 1) (most of the time, U/I and V/I were found negligible).

The polarimeter was calibrated by illuminating the aperture with monochromatic collimated light of known polarization using a fiber-coupled LED, a collimator and a Glan-Thompson calcite polarizer (Thorlabs GTH10M). In addition, an achromatic quarter wave plate (Thorlabs AQWP05M-600) was used for the calibration of the circular polarization. The accuracy of the measurement of Q/I is below ±0.1%. The zero-offset of the instrument, computed by averaging the polarization measured by rotating a half wave plate every 10° from 0 to 360° in front of the aperture, is of 0.001 % which is negligible.

During a measurement sequence, the arm holding the collimated light source is moved from 1.5 to 30° by steps of 1° or more, every 5 seconds. The Stokes polarimeter analyzes the scattered light at 0° emergence continuously, every 60 ms. For each incidence angle, the averages of 4 seconds of measurements of DC and reduced Stokes parameters are computed, together with their standard deviations. A first series of measurements is taken by moving the arm on the right side, from i = 1.5° to 30° and a second series from -1.5° to -30° on the left side, in the principal plane. Because one expects the BRDF to be symmetric around the 0° emergence direction, any asymmetry found can be attributed to the tilt of the sample and/or to a misalignment of the arm or polarimeter. To take into account this source of error, we compute the average of the measurements at +i and -i, together with its standard deviation. To measure the phase curves of the ice samples, the light arm was moved at 1.5° and from 2 to 17° by steps of 1°, and then at 20°, 25° and 30°, lasting only 7 minutes in total. This quick measurement time of the phase curve of all the Stokes parameters is the major advantage of POLICES compared other previous or current instruments, enabling measurements of fresh ice samples and the monitoring their temporal evolution.

The motor of the goniometer arm and the Stokes polarimeter are controlled by a single computer. A homemade program allows to extract the phase curves and calibrate the data.

### 2.1.2 Measurements of photometric standard materials

As mentioned above, the ice samples were illuminated with incident light having a polarization of 2.4 %. We investigated the influence of this polarization on the polarization of the scattered light on standard surfaces of similar albedo than water ice particles: (1) a plate of Spectralon 99% from Labsphere Inc., and (2) a surface made of particles of aluminum oxide ($Al_2O_3$) industrial abrasive with diameter of 3 µm (WCA3 from Micro Abrasives Corporation, Westfield, Massachusetts, USA). Figure 2a,b shows the polarimetric phase curves of these two surfaces, measured when illuminated with a monochromatic light at 530 nm having a polarization of 2.4 % (same as for the ice samples) or 0.05 % (with a depolarizer in the incident light beam). The reduction of the incident polarization results in an average shift of the phase curve of +0.08 % for the Spectralon and +0.12 % for the aluminum oxide particles. We hypothesized that this shift would be similar for surfaces of water ice particles of similar albedo, so we corrected all the measured values of Q/I for the ices by a correction factor of +0.10±0.02 %.

The polarimetric phase curve of 3µm-diameter aluminum oxide particles was also measured by Geake & Geake (1990), Shkuratov et al. (2002) and Ovcharenko et al. (2006). Figure 2 shows that the shape of the phase curve and the absolute values of polarization measured with POLICES at 530 nm are similar to those measured with the polarimeter Minipol at 535 nm by Geake & Geake (1990). Ovcharenko et al. (2006) presented polarimetric phase curves of aluminum oxide WCA3 measured only at 630 nm using the Kharkov photopolarimeter. Figure 2b shows that the polarimetric phase curve measured with POLICES at 625 nm has higher absolute values by up to 0.1% at small phase angles compared to the one measured with the Kharkov's instrument. This may be due to differences in the samples (especially their packing density) or to intrinsic differences in the calibration of the setups. Except this slight difference, the measurements with POLICES agree relatively well with those performed on other instruments. We stress here the need for the measurement of common standard surfaces, such as Spectralon, on the various existing polarimetric setups. To this purpose, we provide in supporting information Data Set S1 the photometric and polarimetric phase curves of Spectralon at 530, 625 and 810 nm measured with POLICES.

### 2.2 Ice samples

In order to investigate the influence of the size/shape of icy particles on their polarization phase curves, we prepared and measured four types of samples, all made of pure crystalline water ice (see Figure 3). The particles shapes and size distributions were measured by Scanning Electron Microscopy (SEM) and/or by optical microscopy, as seen on Figure 3c,f,i,l. An Optical Coherence Tomography (OCT) instrument (Thorlab's Ganymede OCT) was used to perform an interferometric analysis of the light backscattered by the surfaces. Three-dimensional structures of the surfaces and sub-surfaces were obtained (6 µm/pixel) and the roughness was computed, as shown on Figure 3b,e,h,k. The physical properties of the samples are summarized in Table 1.

Spherical ice particles with a diameter of 4.5 ± 2.5 µm, named "spherical S" (or SPIPA-A), were synthesized by freezing a nebula of water droplets produced by an ultrasonic inhalator (see Jost et al., 2016). The frozen nebula was directly deposited onto a cold aluminum plate in contact with a liquid-nitrogen-cooled copper plate under it, producing a smooth and homogeneous surface made of ice spheres, about 2 mm thick (identical to "Method 2" in Jost et al., 2016).

Spherical ice particles with diameters of 70 ± 30 µm, named "spherical L" (or SPIPA-B), were produced by spraying liquid water droplets in a large volume of liquid nitrogen (see Yoldi et al., 2015 or Poch et al., 2016 for details). The particles were deposited in a cylindrical sample holder, about 1 cm-thick, by sieving with a 800 μm sieve to produce a homogeneous layer, free of large agglomerates. The surface was then flattened by cutting the surplus with a cold spatula, without compression.

Crushed ice particles smaller than 400 μm, named "crushed L", were produced by crushing ice cubes using a commercial ice crusher, followed by sieving of the resulting particles. All this preparation was performed in a freezer, at temperatures lower than 238 K, and by mixing the cubes and crushed ice with liquid nitrogen. The particles observed by optical microscopy (Figure 3i) are translucent with angular shapes and a large fraction of them is elongated, with sizes ranging from 200 to 250 μm long and 75 to 150 μm wide. These particles are surrounded by brighter fragments unresolved with the microscope, possibly only few micrometers large. These particles were deposited in a cylindrical sample holder in the same way already described for the "spherical L" particles.

Finally, we prepared frost, simply formed by condensation of water vapor on surface pre-cooled with liquid nitrogen, without manipulation. The frost was left to grow progressively while the phase curves were measured continuously. Two samples were prepared: one with frost forming on a dark aluminum plate ("frost on dark"), and the other forming on the bright surface of a "spherical L" sample ("frost on spherical L"). As the frost grew, the particles size/shape and the surfaces thickness/structure changed; this evolution was monitored using an optical microscope (Figure 3l,m,n). The first particles of frost, formed within the first 3 min following the exposure of the cold surfaces to the atmosphere, seem to be made of roundish particles or aggregates smaller than 10 μm, at the limit of resolution (Figure 3l). They grow with time, forming aggregates on which dendritic needles, tenths of micrometers wide and up to 1-4 mm long, finally form (Figure 3k,m).

The reflectances of the "spherical S", "spherical L" and "crushed L" samples were measured with the POLICES setup just after their preparation, and several measurement sequences were run during about one hour. During this time, the temperature of the samples (initially around 173 K) increased and the particles at their surfaces, in contact with relatively warm air, metamorphosed slowly. The time-series of measurements enable us to follow the temporal evolution of the phase curves as the ice particles metamorphose.

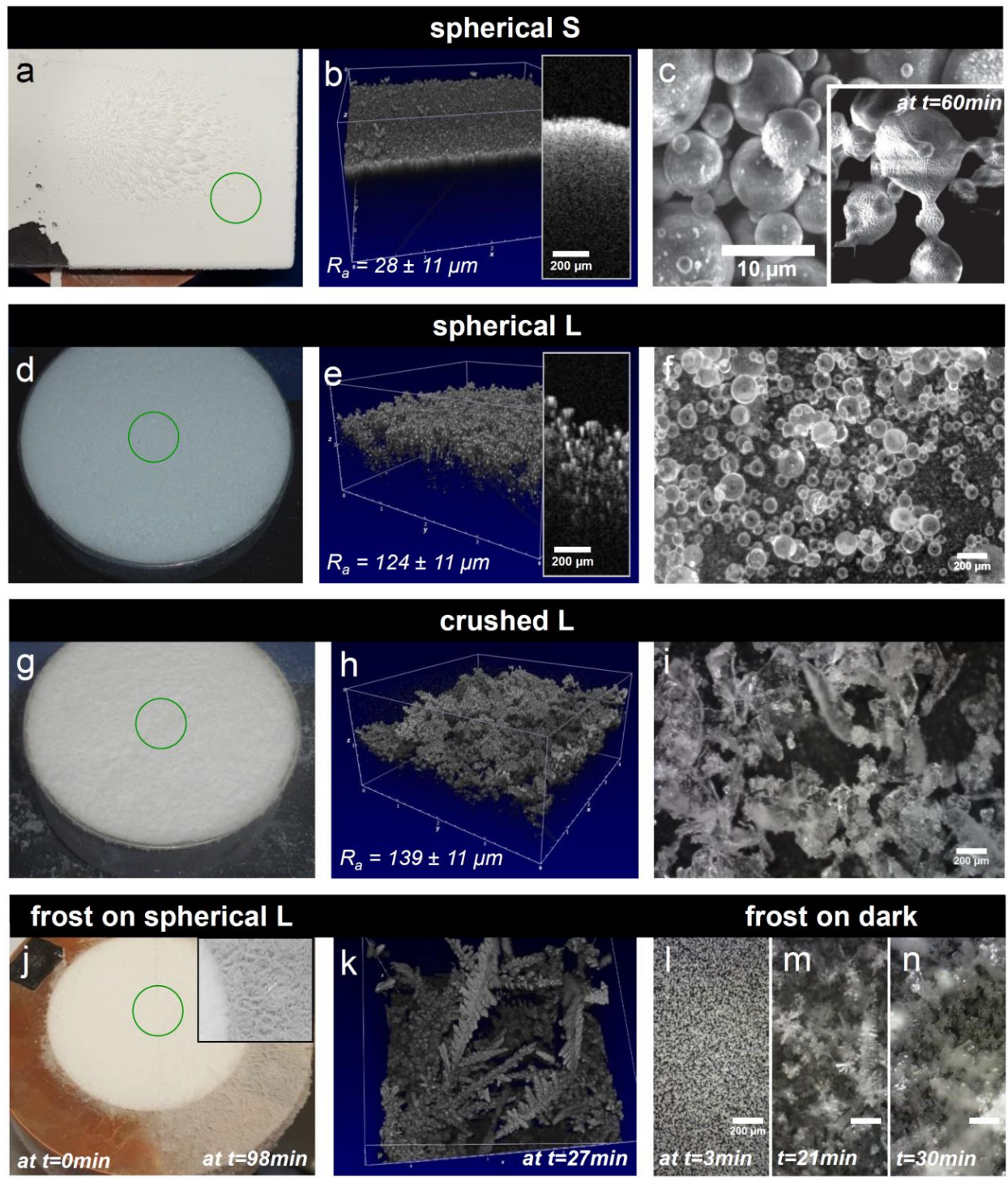

**Figure 3.** Images of the samples: particulate water ice surfaces made of "spherical S" particles of diameter 4.5±2.5 μm **(a,b,c)**, "spherical L" particles of diameter 67±31 μm **(d,e,f)**, "crushed L" particles smaller than 400 μm **(g,h,i)**, "frost on spherical L" **(j)** and "frost on dark" with its temporal evolution **(k,l,m,n)**, from unresolved aggregates or particles smaller than 10 μm at t=3min, to dendrites (10-20 μm wide and 400 μm to 3 mm long) at t=21min, up to sintered frost (larger ice crystals 60-150 μm wide) at t=30min (more images in Figure S1). Figures 3a,d,g,j show

pictures of the samples where the green circle of 15 mm diameter shows the position of the light spot during the polarimetric and OCT measurements. Figures 3b,e,h,k show the 3D structures of the surfaces (4×4mm) with their computed arithmetic average roughness $R_a$ obtained by OCT (arithmetic average height of surface variations from the mean line measured within a sampling length of 4 mm). Figures 3b,e also show OCT depth profiles allowing to compare the light backscattering properties of spherical S and L surfaces. Figure 3c shows SEM images and Figure 3f,i,l,m,n show optical microscopy images of the particles.

**Table 1.** *Summary of the Physical Properties of the Ice Samples at t=0 (t=28min for the frost) and of the Polarimetric Parameters of their Phase Curves shown in Figure 4b.*

| Sample name | Physical properties | | | Polarimetric parameters | | | |
|---|---|---|---|---|---|---|---|
| | **Particles shape** | **Particles size** [a] **(μm)** | **Layer porosity** | **$P_{min}$ (%) (± 0.03)** | **$α_{min}$ (°) (±1)** | **$α_{inv}$ (°) (±1)** | **Slope h** [c] **($10^{-3}$/°) (±0.01)** |
| **Spherical S (SPIPA-A)** | spheres | 4.5 ± 2.5 | 0.70 | -0.47 | 5 | 19.3 | 0.31 |
| **Spherical L (SPIPA-B)** | spheres | 70 ± 30 | 0.45 | -0.14 | 5 | 10.7 | 0.28 |
| **Crushed L** | angular grains | < 400 | unknown | -0.17 | 5 | 11.6 | 0.22 |
| **Frost on spherical L at t=28min** | spheres [b] | < 10 [b] | unknown | -0.49 | 5 | 10 ± 2 | 0.40 |

[a] The values given in this table, and more generally in this paper, are diameters (for spherical particles) or larger dimension (for crushed particles) of the particles.
[b] We did not directly measure the particles' size and shape for this sample, but results from Mie theory modeling suggest the particles were spherical in shape and smaller than 10 μm in diameter.
[c] h is the slope of the polarimetric phase curves obtained by a linear fit of the data points measured from 12 to 30°.

## 3 Results

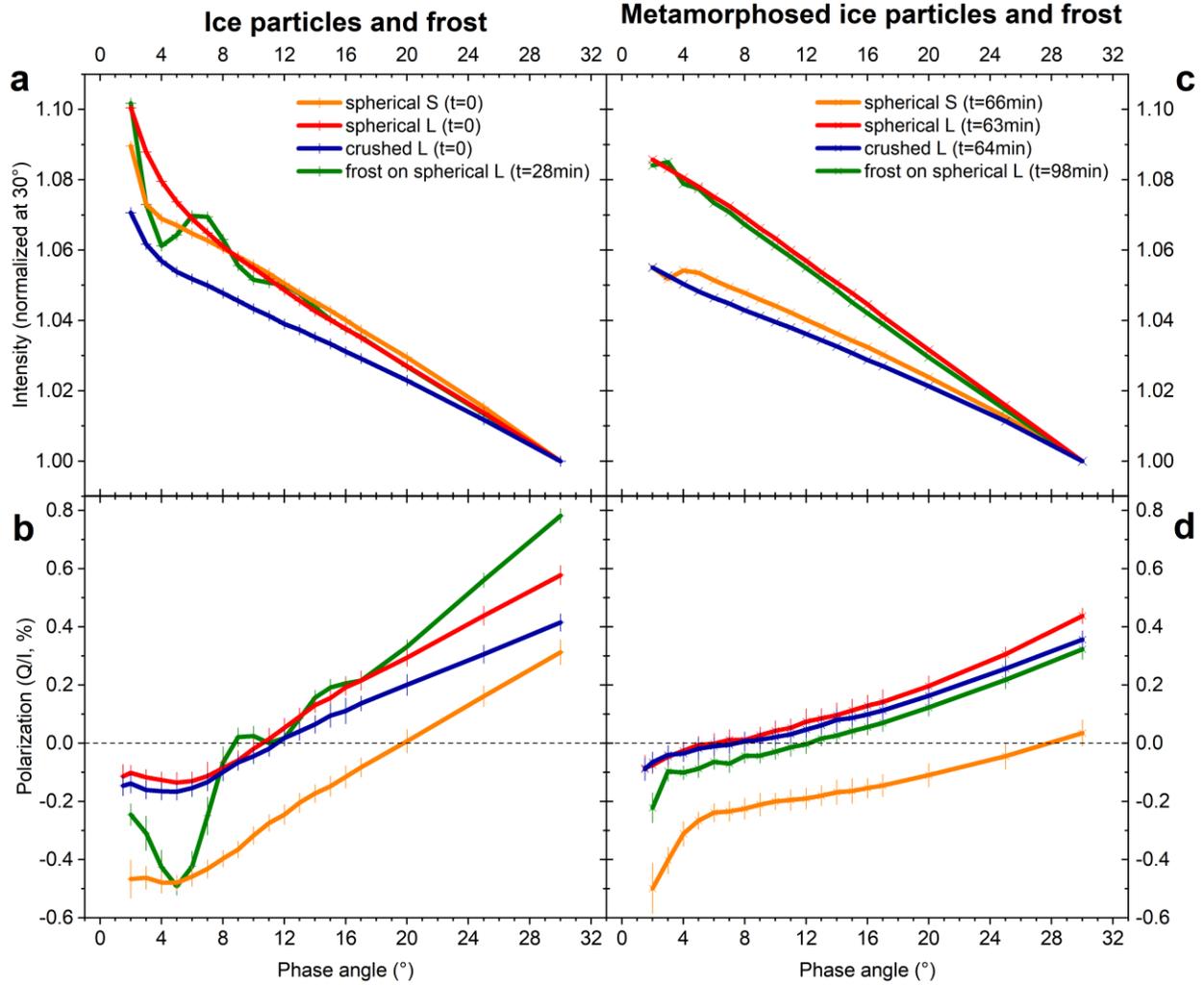

**Figure 4.** Photometric **(a,c)** and polarimetric **(b,d)** phase curves of the fresh ice samples measured immediately after their preparation **(a,b)** and after more than 60 minutes **(c,d)**. The ice surfaces sintered with time, resulting in a vanishing of the non-linear surge of brightness at small phase angles and a shift toward smaller phase angles of the NPB's minimum. The full temporal evolution of each sample is shown in Figure S2 and in the parameter plot of Figure 7. The polarimetric parameters of the phase curves in (b) are shown in Table 1.

### 3.1 Phase curves of the initial fresh ice surfaces

These phase curves shown in Figure 4a,b constitute reference data of well-controlled ice particles size and shape, before their metamorphism.

### 3.1.1. Photometric phase curves

Figure 4a shows that the photometric phase curves of all the samples exhibit a non-linear surge of brightness at phase angles smaller than 5°, the so-called opposition effect. Although POLICES does not allow to measure the whole feature, the surge of brightness from 5° to 2° is clearly different among the samples. For the surfaces made of spherical L and crushed L particles, the rise of brightness is more progressive than for the spherical S particles and the frost, i.e. the width of the opposition peak is larger. This is probably due to the surface roughness of the samples made of spherical L and crushed L particles, higher by a factor of 4 to 5 compared to the surface made of spherical S particles, as measured by OCT (see Figure 3b,e,h). This observation is consistent with the one of Jost et al. (2016) who observed that a rough surface of aluminum oxide particles shows a broader opposition peak than a flat surface at small incidence angles (see Figure 8 in Jost et al., 2016). The main parameter contributing to the width of the opposition peak thus seems to be the macroscopic surface roughness.

Figure 4a also shows that from 30° to 10° phase angles, the relative increase of brightness follows a linear trend with a comparable slope for the spherical S, L and frost surfaces, but a lower slope for the crushed L surface.

The photometric phase curve of the frost at t=0 is characterized by oscillations of increasing amplitudes from about 16° to 2°, with two local maxima at phase angles ~6.5° and ~12°. These are typical signatures of resonances, whose origin is discussed in section 5.1. This sample exhibits the steepest surge of brightness from 3 to 2°.

### 3.1.2. Polarimetric phase curves

The polarimetric phase curves of the fresh surfaces are shown in Figure 4b. It is remarkable to note that the curves of surfaces made of spherical L, spherical S and crushed L particles have a similar shape: a nearly linear trend from 30° to 10° and a parabola-shaped NPB with a minimum at about 5° phase angle.

In the negative branch, the minima of linear polarization range from -0.1% to -0.5%, the surface made of spherical S particles showing the highest absolute degree of polarization.

Interestingly, the surface made of angular and crushed L particles exhibits a negative branch very similar (in shape and intensity) to the one of the spherical L particles. The only slight difference is the lower slope of the phase curve of the crushed L surface. Therefore, the shape of individual ice particles (spherical or crushed) is of little influence on the shape of polarimetric phase curves, which is mostly controlled by other physical parameters of the surfaces.

As for its photometric phase curve, the polarimetric phase curve of the frost deposited on spherical L particles (at t=0) is significantly different from those of the other surfaces, with oscillations from about 16° to 2°.

### 3.2 Phase curves after metamorphization of the ice surfaces

After these initial measurements on the fresh ice samples, follow-up measurements were performed continuously during more than one hour. The ice particles at the surface of the samples, in contact with relatively warmer air, undergo structural changes during this time. Solid bonds of ice are formed between neighboring ice particles via mass transport events (solid state diffusion and/or sublimation and condensation) driven by the reduction of the total surface energy of the system. As a result, the ice particles change of shape and the inter-particulate bonding results in a hardening of the upper surface of the ice samples. This metamorphism of ice particles is referred to as "sintering" (Blackford, 2007). Images of this phenomenon have been obtained for the spherical S particles, as shown in Figure 3c. After sintering, the ice particles exhibit an increase of their size (from $4.5\pm2.5$ to $5.8\pm2.3$ µm), a change of their shape with bonds forming between them and the appearance of ripples-like irregularities carved by local sublimation on their surfaces. The growth and coalescence of particles of frost was also imaged before and after metamorphism, as seen on Figure 3m,n.

The entire temporal evolution of the photometric and polarimetric phase curves of the samples during metamorphism of their ice particles is presented in the Figure S2. The final phase curves measured after more than 1 hour of evolution are shown in Figure 4c,d, where they can be compared with the initial phase curves in Figure 4a,b.

### 3.2.1. Photometric phase curves

As the ice particles sinter, the photometric phase curves get linear and featureless from 2 to 30° phase angles. The non-linear surge of brightness due to the opposition effect completely disappears after 1 hour of evolution (Figure 4c). This evolution is common to all the icy surfaces, irrelevant of their particles size and shape. This is consistent with the observations of Jost et al. (2016) of a decrease of the opposition peak amplitude with time, along with a decrease of reflectance at moderate phase angles on several surfaces made of spherical S particles (Figure 16 of Jost et al., 2016).

However, the evolution of the slope of the photometric phase curve from 5 to 30° differs among the surfaces. Figure 4c shows that the slopes of the phase curves for surfaces made of spherical L particles with and without frost increase with time and reach nearly identical values after 1 hour. On the contrary, the slope of the phase curve for the sample of spherical S particles decreases with time and exhibits a small shoulder at 4°. Furthermore, the slope of the curve of the surface made of crushed L particles remains almost constant during the whole experiment: the phase curve is only affected by the progressive vanishing of the opposition peak (see Figure 4). These disparities indicate that the metamorphism of the ice resulted in different evolution pathways/mechanisms depending on the particles making the surfaces.

### 3.2.2. Polarimetric phase curves

During metamorphism, the polarimetric phase curves of all the surfaces progressively lose their parabola-shaped NPB (Figures 4d and S2). The minimum of polarization ($P_{min}$) is progressively shifted towards smaller phase angles ($\alpha_{min}$), for all the samples. This deformation of the phase curve is more pronounced for the surfaces made of the smallest ice particles (spherical S and frost).

After 1h of metamorphism, the polarimetric phase curves are all featureless, slightly concave and almost parallel to each other from 5 to 30°. They only differ at smaller phase angles. Below 5°, the absolute degree of polarization increases slightly for the surfaces made of spherical and crushed L particles, significantly more for the frost on spherical L particles, and strongly for the surface of spherical S particles. One can note that the polarimetric phase curves of the metamorphosed surfaces contain different signatures at small phase angles indicative of different surface structures. In this respect, the polarimetric phase curves are more informative than the photometric phase curves, which are almost featureless for all surfaces.

The angle of inversion of the polarization ($\alpha_{inv}$) increases of +8° (from 19 to 27°) during the metamorphism of spherical S particles, whereas it only varies of less than ±4° for the other samples (see Figure 7b).

Interestingly, the degree of polarization is almost invariant during metamorphism at some specific phase angles: at ~14° for the spherical S, 1.5-2° and 13-14° for the spherical L, ~14° for the crushed L and ~11° for the frost (see Figure S2).

### 3.3 Phase curves of frost during growth and metamorphism

The phase curves of frost were measured at different times corresponding to different steps of frost formation, growth and metamorphism.

As shown in Figure 5a,b, the first particles of frost formed on the surfaces have their photometric and polarimetric phase curves both characterized by oscillations. As the frost layer grows, the amplitude of the oscillations decreases and their local maxima and minima are shifted towards smaller phase angles, until disappearance of the oscillations (at least from 2 to 30° phase angles). When formed on a dark substrate, the frost exhibits higher degree of polarization and the oscillations have larger amplitudes (see Figure S3a). This is because of the contribution of the substrate to the polarization of the scattered light when the frost is optically thin. The frost on the dark plate was formed under higher relative humidity conditions and air temperature, so it developed much faster than the frost on the spherical L particles. Despite these differences, the phase angles of the oscillations (local minima and period of ~4°), and the general evolution of the shape of the phase curves (from oscillations to flat) are similar for both frost samples.

Figure 5 shows that the oscillations of the photometric and polarimetric phase curves, measured simultaneously during the growth of the frost, were out of phase. The dephasing does not seem to be constant, varying from 1 to 2°, but one can see that the brightness is maximum at phase angles where the polarization varies the most. Figure 5b shows that the derivatives of the polarization phase curves have oscillations with maxima and minima that match relatively well those of the photometric phase curve in Figure 5a.

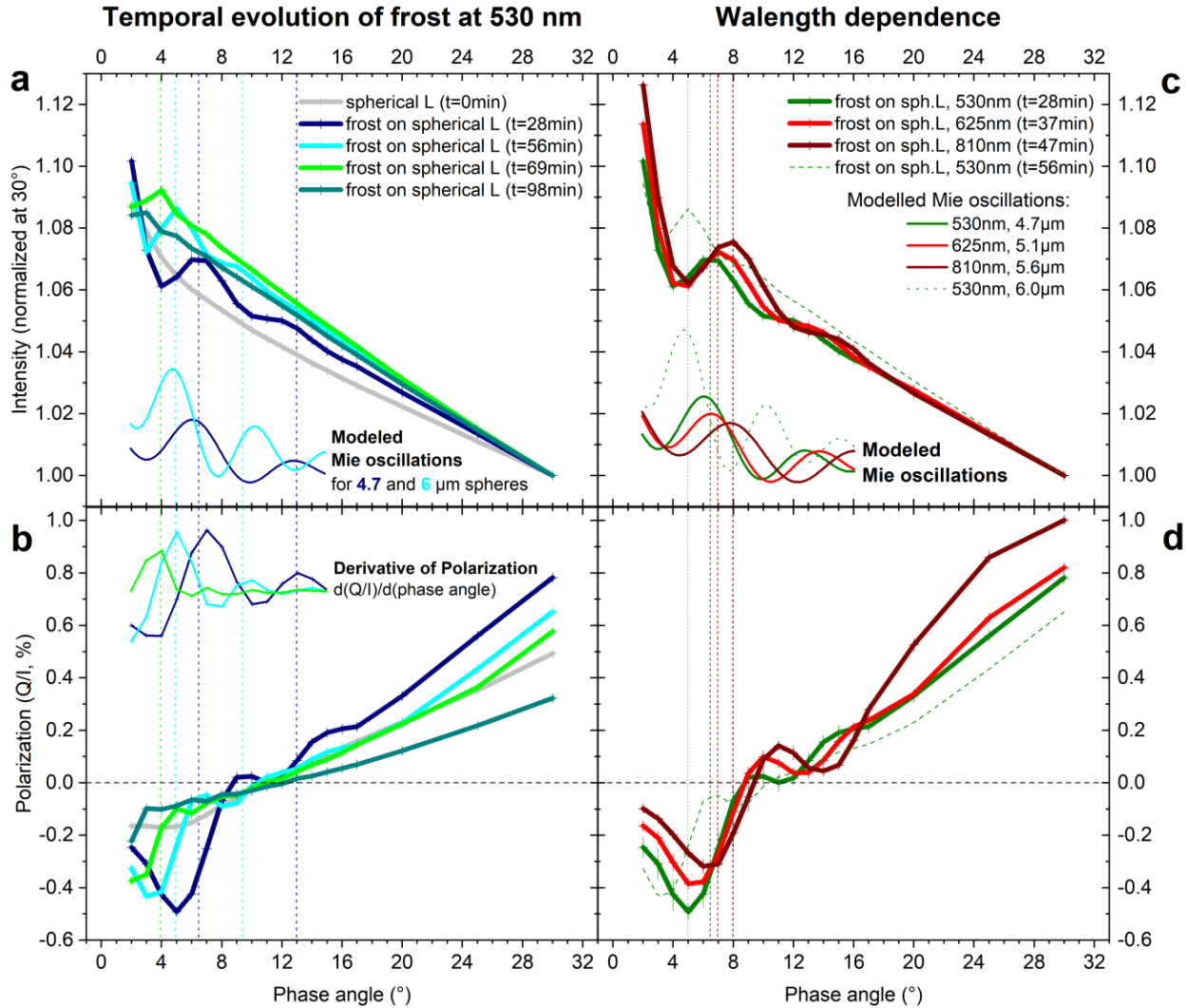

**Figure 5.** Temporal evolution of the photometric **(a,c)** and polarimetric **(b,d)** phase curves of the frost growing by condensation of water vapor on a surface made of spherical L particles and metamorphosing with time, measured at 530 nm **(a,b)** and at 530, 625 and 810 nm **(c,d)**. The lower curves on (a,c) show the results for Mie modeling, matching well the local maxima of brightness (dashed vertical lines). The upper curve on (b) show the computed derivatives of the polarimetric phase curves, matching well the local maxima of brightness. As the frost grows and sinters, the oscillations shift towards smaller phase angles and vanish.

## 4 Discussion

### 4.1 Photometric and polarimetric properties of frost forming on cold surfaces

Figure 5 shows that the early stages of formation and growth of frost on cold surfaces produce spectacular oscillations of brightness and degree of linear polarization with the phase angle. These oscillations are typical for smooth and transparent particles of size parameter (defined as the ratio between the particle size and the wavelength) ranging from 1 to 100, and they are usually called "resonances" (van de Hulst, 1981). At phase angles smaller than 5°, these oscillations are known as the "glory" phenomenon (Laven, 2005b). The mechanisms responsible for the glory and the resonances are not fully understood. They probably result from the interference between light rays reflected and refracted by the particles as well as surface-waves (Bryant & Cox, 1966; van de Hulst, 1947, 1981; Nussenzveig, 1979). Nevertheless, for spherical particles, this interference pattern can be accurately computed using the Mie theory (Laven, 2012).

We used the program "MiePlot", developed by Philipp Laven (http://www.philiplaven.com/mieplot.htm) to perform computations of the Mie scattering by spherical particles of ice, in order to gain more insights into the properties of the frost and their temporal evolution. The maxima and minima observed on the photometric phase curves measured at 530 nm on frost formed on spherical L particles at t=28 min and t=56 min are best fitted by Mie oscillations generated by spherical water ice particles of 4.7±0.2 µm and 6.0±0.3 µm diameter respectively (see Figure 5a and Figure 5c). To confirm that these oscillations follow the wavelength dependence predicted by the Mie theory, we have also measured the phase curves of the frost at two other wavelengths (625 and 810 nm) as it grew from t=28 min to t=56 min. These measurements are shown in Figure 5c,d together with modeled Mie oscillations for particles of increasing diameters, assuming a constant growth from t=28 min to t=56 min. The modeled oscillations nearly perfectly match the observed ones.

The coincidence of the observed oscillations with the Mie computation implies the presence of spherical particles in the frost. Microscopy images of the early frost layer formed on the dark plate (Figure 3l) do not show these perfectly spherical particles, probably because they are under the limit of resolution of the microscope.

The nucleation of frost on cold surfaces has been studied, both experimentally and theoretically and is still an active field of research (Li et al., 2017; Piucco et al., 2008; Wu et al., 2007a; Wu et al., 2007b). Wu et al. (2007a) observed that the first ice embryos are formed by condensation of water vapor into spherical super-cooled liquid droplets on surfaces at temperatures ranging from 267 to 255 K. These droplets grow within few minutes until they finally freeze, having final diameters from 40 to 150 µm under a relative humidity of 35%. Then, frost crystals form and grow on these frozen droplets, forming various structures (flakes, needles, feathers etc.) depending on the temperature and humidity. From the observations of Mie oscillations during our experiments, we can deduce that spherical embryos of frost are formed on surfaces at 173 to 220 K and relative humidity of 25-30%. But they are much smaller than those observed by Wu et al. (2007a), having diameter around 4 to 6 µm as inferred by the Mie modeling. This observation favors the scenario proposed by Li et al. (2017) that for ultra-low surface temperatures, liquid droplets may form from the vapor close to the cold surface before being deposited and frozen on that surface.

Moreover, the secondary maxima of the Mie oscillations should rapidly vanish when the standard deviation of the size distribution gets larger than 15% (see modeling results on Figure S5a). Consequently, the presence of not only one but also several local maxima on each phase curve of early frost (blue curves on Figures 5 and S3) is an indication of the relatively narrow size distribution of the frost embryos. This is probably a consequence of their nearly simultaneous formation and growth.

The angular shift of the oscillations (Figures 5 and S3) is probably due to the growth of the spherical frost embryos with time, as attested by the modeled phase curves presented in Figure S4b. In addition, as the frost grows, the shape and size of the particles are modified: dendrites with trunks and branches of several tenth of micrometers wide, and from 400 µm to 3 mm long developed on the initial embryos (see Figures 3k,n and S1). These changes of particles shapes and size distribution contribute to the vanishing of the oscillations.

When the frost layer is made of a porous network of elongated dendrites (layer 1 to 3 mm thick), it becomes thermally insulating and its upper part reaches temperatures very close to 273 K (Kim et al., 2017). The frost dendrites are transformed in larger ice crystals, more translucent and with angular facets, 60 to 150 µm large (Figure 3n). This very metamorphosed ice surface corresponds to the green phase curve at t=30 min shown in Figure S3a. It is devoid of oscillation and very similar to the phase curves of the other metamorphosed ice surfaces in Figure 4c,d, indicating that they probably reached similar temperatures and are also made of translucent water crystals.

The apparent relation between the oscillations of the brightness and polarimetric phase curves remains unexplained (Figure 5b). Further modeling work is needed to understand these observations.

In the past, Dougherty and Geake (1994) measured the polarization of frost formed on a black-painted plate cooled with liquid nitrogen, with very similar values of $P_{min}$ and $\alpha_{inv}$ than our "frost on dark". They did not report resonances, although they noted a tendency for the shape of the NPB to be variable and complicated. Their longer acquisition time (60 min versus 3.5-7 min) probably prevented them to observe the oscillations.

### 4.2 Influences of the ice particles properties on the polarimetric phase curves

The data set presented in Figure 4b provides reference polarimetric phase curves for surfaces made of well-characterized ice particles. In this section, we discuss how the ice particles physical properties influence the NPB of their phase curves.

As we mentioned above, the observation of Mie oscillations requires the presence of transparent spherical particles having a diameter below 50 µm and a relatively narrow size distribution. The absence of oscillation on the phase curves of spherical S and L particles may be due to their internal structure (opacity) and/or to their size (spherical L particle may be too large to exhibit resonances) and/or to their broad size distributions ($\sigma \approx 50\%$) relative to the frost embryos ($\sigma < 15\%$). The spherical L and S particles are formed by freezing liquid droplets of water. Water freezes from the exterior to the interior of the droplets in a process that may create stresses due to the expansion of the ice, resulting in the formation of internal cracks, faults or other optical defects that prevent the internal propagation of the light rays producing resonances. On the Figure 3f, the spherical L particles of ice appear to be rather transparent, but some internal structure can be seen, which may prevent resonances; this may be also true at smaller scale for the spherical S particles (too small to image their internal structure using optical microscopy). Conversely, the spherical nuclei of frost are formed by direct condensation of water vapor, in a gentler process, probably allowing the formation of perfectly transparent particles, without internal defects, exhibiting resonances.

We note here that in a previous study of the photometry of spherical S particles using the PHIRE-2 goniometer, we reported the observation of weak maximum of brightness around 2 to 5° phase angle, which disappeared after few hours (Jost et al., 2016). This feature, that we attributed to the glory effect, is clearly absent from our data (Figure 4a). We cannot exclude that the glory observed on spherical S particles in 2016 might be due to frost embryos formed on the sample, whose ability to produce this effect was unsuspected to us at that time. This demonstrates the difficulty to work with water ice particles and the need of constant improvements in the characterization and the measurement conditions of the samples.

The particle size influences significantly the degree of polarization. The larger particles (spherical and crushed L) have shallower NPBs than the smaller particles (spherical S and frost). A decrease of the ice particles diameter by a factor of 15 between spherical L and spherical S samples induces a decrease of $P_{min}$ (increase of $|P_{min}|$) by a factor of ~3. Moreover, spherical S and fresh frost particles have the same value of $P_{min}$ (-0.5%) and similar average particles sizes: respectively 4.5±2.5 µm as measured by SEM and 4.7±0.2 µm as modeled using "MiePlot". However, it is less clear why the crushed L sample has a very similar value of $P_{min}$ than the spherical L sample ($\approx$ -0.15%). The exact size distribution of the crushed L particles is unknown but the microscopy images show that it appears to be made of a mixture of particles from few micrometers up to 250 µm. Thus, its average grain size might be comparable to that of the spherical L particles (70 ± 30 µm). All these observations indicate that $P_{min}$ seems correlated to the average size of pure particles of ice. This is in agreement with previous works by Geake & Geake (1990) and Nelson et al. (2018) with aluminum oxide particles, who described a continuous change of the shape and deepening of the NPB with decreasing particle size to wavelength ratio ($d/\lambda$).

It is striking that the spherical S particles have the largest value of $\alpha_{inv}$ compared to the other samples. This is likely due to a higher multiple scattering of the light by this surface. Figure 3b,e shows OCT data, indicating the higher density of light backscattering events in a fresh surface of spherical S particles compared to a surface made of spherical L particles. The surface of spherical S particles offers, in a given volume, more interfaces for the light rays to interact with, allowing more multiple scattering. Mishchenko et al. (2009) demonstrated that increasing the number of particles in the modeled volume results in increasing the amount of multiple scattering within the surface layer, producing a narrower opposition peak and a deeper NPB via the coherent backscattering (CB) phenomenon. CB is due to the constructive interference occurring around 0° phase angle between two rays of light scattered multiple times within the surface material and following the same path in opposite directions. Furthermore, an increase of the amount of multiple scattering does not change $\alpha_{min}$ but can shift $\alpha_{inv}$ to larger values (see Figures 3 and 4 in Mishchenko et al., 2009), as seen with spherical S versus L particles (Figure 4b).

Finally, another polarimetric parameter that we did not discussed here is the slope from 5 to 30° (see values in Table 1). Previous works have shown that this parameter appears to be mainly correlated to the albedo, which was not measured in the present work (Geake & Dollfus, 1986 and references herein, Dougherty & Geake, 1994).

4.3 Evolutions of the phase curves during the metamorphism of the ice particles

Mishchenko et al. (2009) and Muinonen et al. (2012) demonstrated via different numerical modeling efforts that the phenomena of the opposition effect and the NPB are due to a combination of the shadow-hiding and the CB mechanisms. During metamorphism, the particles sinter, resulting in an increase of their size and a decrease of the number of small particles that are more incline to coalesce. Consequently, the CB and the shadow-hiding effects may become less and less pronounced, causing the simultaneous vanishing of the opposition peak and of the NPB.

We note that the phase curves of metamorphosed ice samples (Figure 4d) are quite similar to the one of the Spectralon surface shown in Figure 2a. Interestingly, this surface, manufactured by Labsphere, Inc., is also produced by sintering of a powder of polytetrafluoroethylene particles (Helmlinger & Arrecchi, 2012). SEM images of the surface of Spectralon show an amorphous structure devoid of individual particles, connected by sub-micrometer-sized filaments (see Figure 16b in Parretta & Addonizio, 2015). Sintered ice surfaces might share some similarities with this structure, being made of a crust-like layer of ice where individual particles are absent but void volume is still large.

While $P_{min}$ increases and $\alpha_{inv}$ hardly varies during metamorphism of most of the ice samples, the one made of spherical S particles exhibits a significantly different evolution: $P_{min}$ is stable and $\alpha_{inv}$ increases very significantly (Figures 4 and 7b). The stability of $P_{min}$ may be explained by the simultaneous increase of two populations of light scatterers, larger and smaller than the initial ice particles, maintaining the stability of the average particle size. The first population could be the larger particles formed by sintering and the second one could be the ripples-like irregularities and/or the bridges between particles, which constitute new scatterers of much smaller sizes (about 0.4 to 1 µm as seen in Figure 3c) than the initial ice particles. These smaller new scatterers may be responsible for an increase of multiple scattering, explaining the increase of $\alpha_{inv}$.

4.4 Application to the characterization of the surfaces of icy satellites

The measurements presented in this study demonstrate the high sensitivity of visible polarimetry to the grain size and degree of sintering of surfaces made of water ice particles. Consequently, visible polarimetry is of obvious interest to characterize icy planetary surfaces, either on Earth, Solar System bodies or exoplanets.

Ground based polarimetric observations of the icy satellites of Jupiter, Saturn and Uranus have been performed since the 1960s (Mishchenko et al., 2010; Rosenbush et al., 2015; Zaitsev et al., 2013). Figure 6 presents the average disk-integrated polarimetric phase curves obtained from the fits of the polarization data for Europa and some Saturn's moons, as shown in Rosenbush et al. (2015).

The polarization phase curves of Io, Europa and Ganymede are characterized by a wide NPB from about 2 to 9° and a narrow asymmetric minimum of negative polarization at around 0.2-0.4°, called the "polarization opposition effect" (POE). The minimum degree of linear polarization in the NPB ($P_{min\ NPB}$) is around -0.2% to -0.3% and in the POE ($P_{min\ POE}$) it goes down to about -0.4% to -0.5%. The polarimetric phase curve exhibits a significant scatter of the data points of about 0.2% (see data points for Europa in Figure 6), which may be due to longitudinal variegation of the satellites' surface properties (Rosenbush et al., 2015). The average trend of the NPB appears relatively flat over wide intervals of phase angle, and the switch of the polarization sign is abrupt, occurring in a very narrow interval of phase angles between 8.6 to 9.8° depending on the satellite (see Fig. 3.9 in Mischenko et al., 2010). Less observational data are available for the satellites of Saturn, but they clearly exhibit deeper minima of polarization (higher $|P_{min}|$) than the Galilean satellites. For example, the icy moons Rhea and Dione exhibit a polarization of -0.9 to -0.6%. The major satellites of Uranus show even deeper minima of polarization from -1.1 to -1.4%. Moreover, the observations show that in general as $P_{min}$ decreases $\alpha_{min}$ is shifted towards smaller phase angles. The origin of these polarimetric properties are still unknown.

Our polarimetric measurements of well-characterized pure water ice particulate surfaces help to understand these observations. Figure 6 shows a comparison of the laboratory measurements with the average disk-integrated polarimetric phase curves of some of the satellites having the purest water ice surfaces and highest albedo: Europa, Enceladus and Rhea, having geometric albedos of 1.02, 1.38 and 0.95 respectively (Verbiscer et al., 2013). We limit our comparison to these bodies because their surfaces contain a relatively low amount of dark non-icy materials, which can significantly affect the degree of polarization. As an example of this effect, we also plot the polarimetric phase curve of the trailing hemisphere of Iapetus, having an albedo of 0.55, to compare it with frost grown of a dark surface. In addition, Figure 7 provides a graphical way to compare the polarimetric properties of icy satellites with those of the samples.

Figure 6a shows that the average phase curve of Europa displays a similar value and position of polarization minimum ($P_{min \, NPB}$) than the surfaces of spherical L (70±30 µm) and crushed L (< 400 µm) ice particles. However, the switch of polarization sign occurs at lower phase angle for Europa ($\alpha_{inv} = 8.7\pm0.1°$) than for the spherical/crushed L ice particles ($\alpha_{inv} = 11\pm1°$). Interestingly, when these ice particles sinter, their inversion angle becomes closer to the one of Europa (changing from 11 to 7°) but the shape of the curve gets flatter (Figure 6c). These data appear compatible with a scenario where the surface of Europa is dominated by ice particles of ~40-400 µm diameter having different degrees of sintering. The analysis of the water absorption bands from mid-infrared spectroscopy observations of Europa have led to the same grain size range, from 40 to 400 µm (Dalton et al., 2012; Hansen and McCord, 2004; Hansen, 2005; Ligier et al., 2016).

The abrupt change of sign of the polarization (Fig. 3.9 in Mischenko et al., 2010) remains unexplained. Indeed, all our measurements are consistent with a progressive decrease of the absolute level of polarization near the inversion angle. This peculiar change of sign is thus likely to be due to a disk-integration effect (combining various surface areas having different polarimetric properties and orientations) and/or to the influence of the Jupitershine as proposed by Mischenko et al. (2010) and Rosenbush et al. (2015). Additionally, although the range from 0 to 1.5° was not accessible to our measurements, it is interesting to note that sintered porous samples such as frost on spherical L at t=98min exhibits a sharp decrease of polarization with values and angular positions not far from the ones observed for Europa's POE (Figure 6c).

Concerning the Saturn's moons Enceladus, Rhea and the trailing hemisphere of Iapetus, Figure 6b shows that their phase curves are best fitted by relatively fresh frost. Their low values of $P_{min}$ are compatible with ice particles of few micrometers in diameter, but their shape is very different from that of the spherical S particles (4.5±2.5 µm) and more consistent with the narrower parabola-like negative branches of fresh frosts. Mie modeling of the phase curves of frost on spherical L at t=28 min and t=56 min gives diameter of ice spheres of 4.7±0.2 μm and 6.0±0.3 μm respectively (see Figure 5a and Figure 5c). These values suggest that the surface of Enceladus is covered with ice particles of about 5 μm diameter in average and Rhea of even smaller particles.

The estimation of the ice particle size from mid-infrared spectroscopy observations of Saturn's moons give different results depending on the method used to interpret the data. Mie calculations also give particle sizes ranging from sub-micrometer-sized to 5 μm for Mimas, Tethys, Rhea, Enceladus, Dione and the icy hemisphere of Iapetus, compared to larger sizes, from 10 to several 100 μm, for the Galilean satellites (Stephan et al., 2016; Hansen, 2005). Modeling of the phase angle variations of the spectra of some areas of Rhea, Dione and Tethys using the Multi-Sphere T-Matrix code also suggests icy particles size of 2 μm in diameter (Pitman et al., 2017). Conversely, for Enceladus, Verbiscer et al. (2006) found grain sizes of 40-60 μm in diameters, Jaumann et al. (2008) found a global particle size distribution peaking at 20 μm and Filacchione et al. (2012) obtained 63 μm, while Scipioni et al. (2017) found the spectra compatible with the presence of sub-micrometer-sized ice particles. Using Hapke modeling, Filacchione et al. (2012) found a mean diameter of 38 μm for Rhea, a result about a factor of ten larger than the other estimates, like for Enceladus. Hansen (2009) pointed out discrepancies in modeled grain sizes between Mie and Hapke modeling, and recommended the use of Mie calculations for analyzing unknown spectra. Here we note that the polarimetric data seem to give grain sizes similar to Mie modeling of infrared spectroscopic data. The trailing hemisphere of Iapetus, although rich in water ice, shows a lower geometric albedo than Enceladus and Rhea because its surface contains some amounts of dark dust (Clark et al., 2012). It is interesting to note that the polarimetric phase curve of frost grown on a dark surface at t=14 min seems relatively similar to that of Iapetus trailing hemisphere (Figure 6b). This also illustrates the fact that $P_{min}$ can only be a reliable proxy of the grain size when comparing surfaces of similar albedo and composition.

Finally, clearly none of the phase curves of metamorphosed ice samples looks similar to those of the Saturn's moons (Figure 6c), indicating less consolidated, more pristine, icy surfaces or evolution processes very different from the metamorphism observed in our laboratory conditions.

The photometric phase curves of the icy satellites also exhibit differences that can be due to dissimilar surface particles structure and texture. From 1 to 50° phase angles, Europa's photometric phase curve is flatter than Enceladus and Rhea (Buratti et al., 2008; Buratti 1995; Grundy et al., 2007; Domingue et al., 1997). In addition, from 0 to 1° phase angles, the amount by which the reflectance increases towards opposition is lower for Europa than for Enceladus and Rhea (Verbiscer et al., 2013). Differences of surface roughness, porosity and/or particles size influencing the shadow-hiding and CB mechanisms may explain these observations whose interpretation remains difficult. Our data show that the surge of brightness at the smallest phase angles measured here (between 3 and 2°) is the highest for the smallest ice particles (Figure 4a) and the lowest for the sintered ones (Figure 4c). Consequently, at small phase angles, both the photometric and polarimetric data could be consistent with larger and/or more sintered ice particles on Europa than on Enceladus and Rhea.

To conclude, our data suggest that the change of polarimetric properties of the icy satellites with decreasing distance from the Sun, pointed out by Rosenbush et al. (2015), may be due to increasing sizes and degree of sintering of the ice grains covering their surfaces. Galilean satellites may be covered by relatively coarser grains (~40-400 μm and larger) than the Saturn's moon, made of finer frost particles. This could explain the lower thermal inertia of the Saturn's moons compared to the Galilean satellites (Howett et al., 2010), because surfaces made of smaller grains have lower thermal conductivity (Presley & Christensen, 1997). The possible dissimilarity of grain size between Saturn's and Jupiter's moons may be due to their different surface temperatures, ranging from 80 to 130 K for Europa, but from 50 to 100 K for Enceladus and Rhea (Howett et al., 2010; Spencer et al., 1999). As shown by Clark et al. (1983) and Gundlach et al. (2018), temperatures above 100 K allow the mobility of the ice, resulting in grain growth and sintering, impeded at lower temperatures. Moreover, the resurfacing mechanisms of these satellites may be significantly different. Irradiation by electrons and ions from the magnetosphere is less severe on Saturn's than on Jupiter's satellites (Mauk et al., 2009) and sputtering rate is observed to be correlated with grain size on Europa (Cassidy et al., 2013). Additionally, deposits of endogenous or exogenous particles may also influence grain sizes, such as the micrometer-sized ice particles from the E-ring, which are suspected to coat the Saturn's satellites (Howett et al., 2010; Verbiscer et al., 2007).

As a final point, we note that caution should be taken when comparing laboratory measurements, performed at a single emergence angle, with disk-integrated observations, which contain the contributions of multiple light rays scattered on the planetary hemisphere with emergence angles varying greatly. Indeed, to our knowledge it has not been proved that the degree of polarization at a given phase angle is independent on the emergence or incidence angles. To allow a more robust comparison with observations, it would be necessary to measure bidirectional polarization distribution functions (BPDF) of surface samples and then use these functions to compute the disk-integrated phase curves. Future laboratory works should address this point.

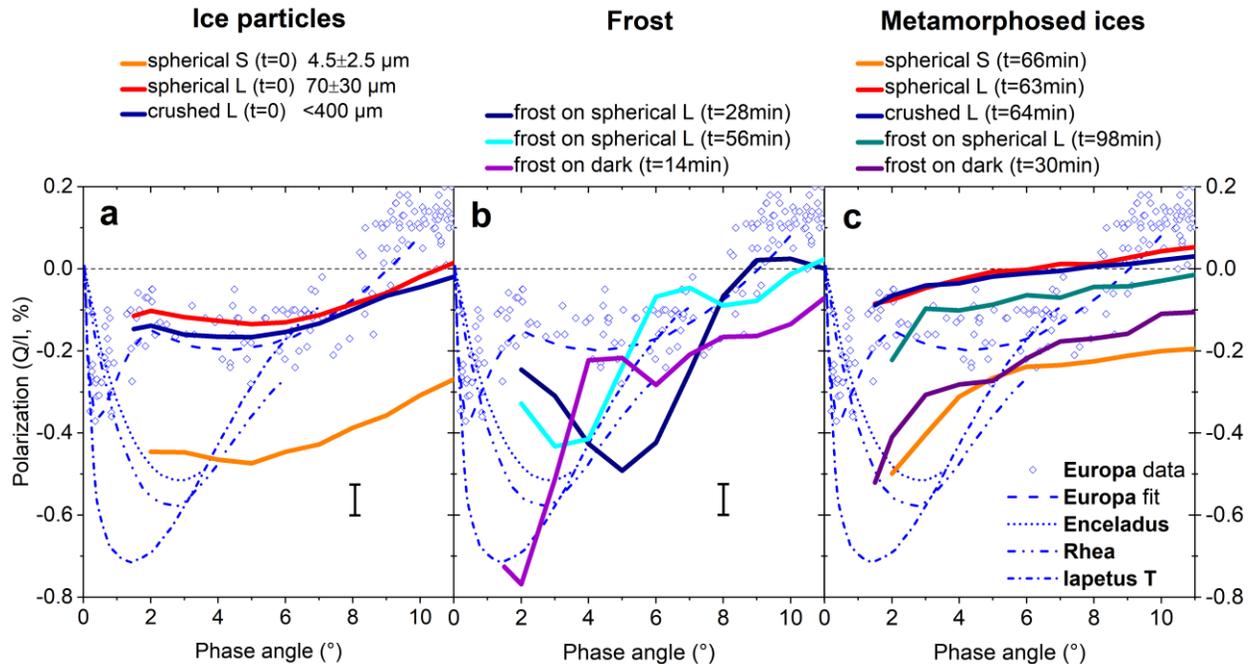

**Figure 6.** Comparison of our laboratory measurements of water ice particles (at 530 nm) with the disk-integrated polarimetric observations of icy satellites Europa, Enceladus and Rhea and Iapetus

trailing hemisphere. Blue diamonds are data points of Europa observations acquired in spectral band V or centered at 550 nm from the database of Zaitsev et al. (2012). Blue curves are average phase curves of the satellites obtained by fitting observations in bands V or R, from Rosenbush et al. (2015). The error bar indicates the typical error of the laboratory measurements. The observations of Europa are consistent with large and/or sintered ice particles, whereas the data from the Saturn's moon are more consistent with finer frost.

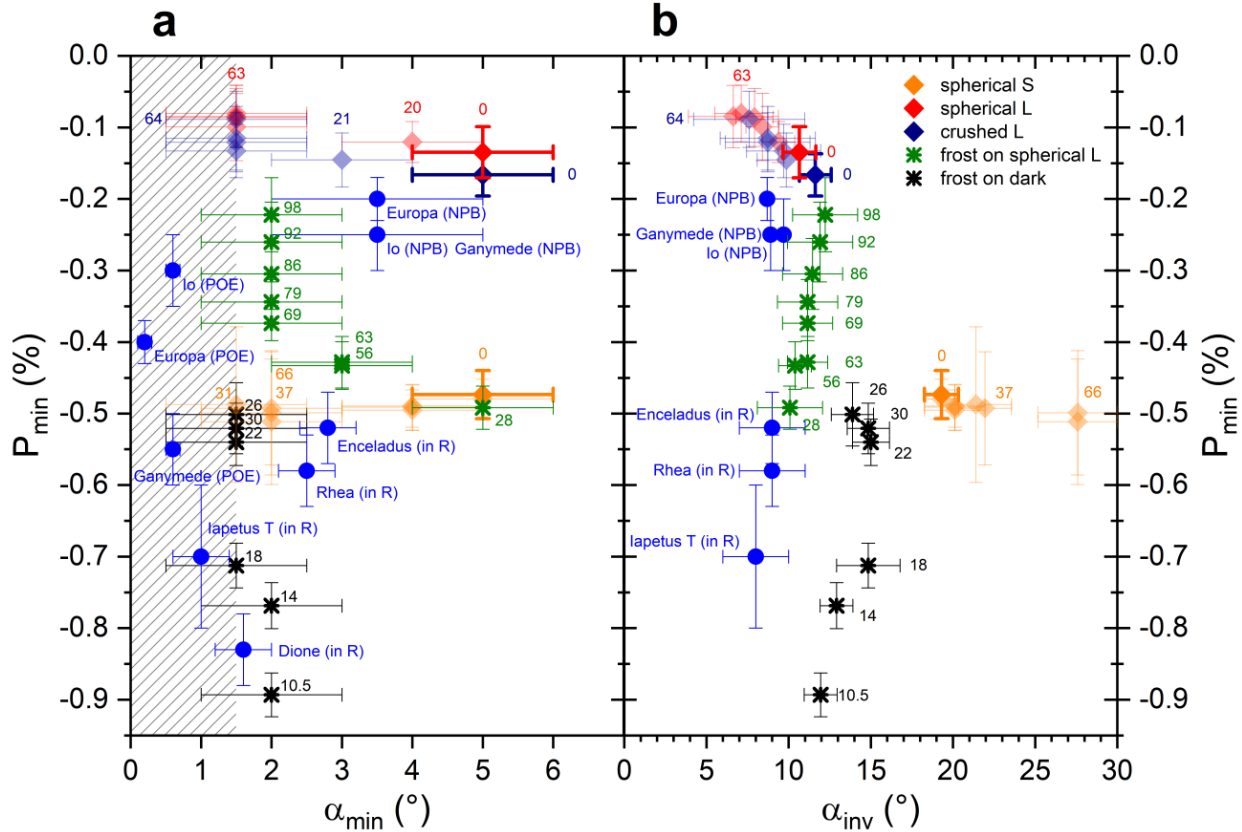

**Figure 7. (a)** Depth ($P_{min}$) versus phase angle at minimum ($\alpha_{min}$) or **(b)** Depth ($P_{min}$) versus width ($\alpha_{inv}$) of the negative part of the polarimetric phase curves of the ice samples shown in Figure 4, Figure S2, Figure S3 and of the disk-integrated polarimetric phase curves of some icy satellites described in Rosenbush et al. (2015). Part of the plot is hatched because POLICES cannot measure phase angles smaller than 1.5°. The number next to each data point gives the time of the measurement in minutes after t=0. The measurements performed at t=0, corresponding to well-characterized samples (Figure 4b), are highlighted in bold diamonds. During time, the ice surfaces underwent thermal sintering. Frost samples are particular cases because the frost also grew with time as water condensed on the cold surfaces. Laboratory and astronomical data are given at 530 nm and V spectral band (550 nm) respectively, except for the Saturn's moons for which data in R spectral band (660 nm) are shown. NPB stands for "negative polarization branch" and POE for "polarization opposition effect". These plots clearly show the increase of $P_{min}$ for increasing particle size and the decrease of $\alpha_{min}$ during sintering. They also show that the phase curves of the Galilean satellites are consistent with large grain sizes or sintered frost whereas those of the Saturn's moons are more consistent with finer frost. All these data are available in supporting information Data Set S2.

## 5 Conclusions and Perspectives

We have measured the photometric and polarimetric phase curves of several surfaces made of well-characterized pure water ice particles (spherical and crushed grains produced from freezing liquid water, frost from condensation of vapor), mainly at 530 nm and from 1.5 to 30° phase angles using the newly developed POLICES setup at the University of Bern.

These measurements reveal **how the parameters of the negative polarization branch are influenced by the properties of the particulate ice surfaces**:

- The results show that the minimum of polarization in the negative branch ($P_{min}$) is the main indicator of the average particle size: the larger by absolute value $P_{min}$, the smaller the particle size.
- We have found that fresh frost formed by condensation of water vapor on cold surfaces (at 173 to 220 K and RH = 25-30%) produces phase curves characterized by oscillations of the degree of polarization by up to a factor of 10 from 0 to 8°. These oscillations are due to resonances, typical for light scattered on smooth and transparent particles of simple shapes. They are well reproduced by computed Mie oscillations, suggesting the presence of spherical frost embryos of few micrometers and narrow size distribution ($\sigma <$ 15%). As the frost layer grows, the angular position of the oscillations is shifted towards smaller phase angles while they progressively vanish due to the mixed influences of (i) the increase of particle sizes, (ii) the broadening of the size distribution and (iii) the change of particles shapes.
- Alternatively, spherical particles of ice formed by rapid freezing of liquid droplets have similar polarimetric phase curves as crushed particles, with a larger parabola-like negative branch without oscillation. This is probably because they contain internal optical defects on which the light scatters, causing multiple scattering and destroying the resonances but producing a shallow parabola-like NPB typical for particulate surfaces.
- Our data also suggest that the angle of inversion of the polarization ($\alpha_{inv}$) increases with the amount of multiple scattering in the surface, controlled by the number of light scatterers by unit of volume.

To summarize, the polarimetric phase curves of surfaces made of water ice particles appear to be mainly influenced by (i) the particle size (through the $d/\lambda$ ratio), (ii) the width of their size distribution, (iii) the intra- and inter-particles structure of the ice layer.

Moreover, we found that **the polarimetric phase curves are extremely sensitive to the degree of sintering of the ice surfaces**:

- Fresh ice samples exhibit a strong non-linear surge of brightness, the opposition peak, together with a parabola-shaped negative polarization branch (NPB). As the ice surfaces undergo thermal metamorphism, these two features vanish simultaneously.
- Independently from their initial state (spherical, crushed or frost particles), all ice surfaces affected by thermal metamorphism see the slope of their polarimetric phase curves from 5 to 30° decreasing towards similar values and their NPB's minimum shifting towards smaller phase angles: $\alpha_{min}$ decreases below 2°. However, sintering leads to different slopes of the photometric phase curves from 5 to 30° and different shapes of the polarimetric phase curves below 5°, depending on the initial samples.

These evolutions observed during sintering are due to the modifications of the ice particles and surface structure driven by the reduction of the surface energy, coalescence of grains forming larger particles, but also local sublimation/recondensation forming smaller light scatterers, which can have diverse influences on the optical properties. Further experimental characterization of samples undergoing sintering and numerical modeling are needed to explain which optical mechanism(s) play a role in these evolutions.

These measurements also constitute references for the direct interpretation of observations through comparisons of data. They shed light on the **surface properties of the icy satellites of the Solar System**:

- Comparisons of the polarimetric phase curves of pure water ice particles with the observations suggest that Europa is possibly covered by relatively coarser (~40-400 μm) and more sintered grains than the surfaces of Enceladus and Rhea, which are more likely covered with finer frost-like particles of few micrometers in average. These conclusions obtained from polarimetric data in the visible domain are in line with some of those obtained from the analysis of the water absorption bands via mid-infrared spectroscopy (Dalton et al., 2012; Hansen and McCord, 2004; Ligier et al., 2016; Stephan et al., 2016; Hansen, 2005; Pitman et al., 2017).

However, the surfaces of these icy satellites is not made of pure water ice and contain some amounts of various mineral and organic compounds (Dalton et al., 2012; Filacchione et al., 2012). Consequently, we foresee to extend this laboratory work to mixtures of water ice with salts, sulphur and other compounds to study their influences on the polarimetric properties of icy surfaces. Future measurements should also study the effect of the porosity of layers of ice particles on the polarimetric phase curves.

Because of its great sensitivity to grain size and degree of sintering of ice particles, **polarimetry could be used to detect hints of ongoing processes on icy planetary surfaces**. The surfaces of the icy satellites of the giant planets are affected by endogenous (tectonic activity, cryovolcanism) as well as exogenous (radiolysis and sputtering due to energetic particles) processes responsible for changing the surface micro-texture globally or locally (Cassidy et al., 2013; Jaumann et al., 2008; Scipioni et al., 2017; Sparks et al., 2017). Spatially resolved polarimetric observations of the icy satellites could help to identify areas affected by these processes. Such observations can be achieved using large ground based telescopes, such as the VLT equipped with SPHERE and its high-angular resolution polarimetric imager ZIMPOL (Schmid et al., 2018). Much higher spatial resolution could be provided by polarimetric instruments onboard spacecrafts. In this respect, it could be interesting to re-analyze the old data gathered by the photopolarimeters onboard Galileo, Pioneer 10 and 11 (Martin et al., 2000), and consider the development of new polarimetric instruments for future missions.

The results of this study could also find applications in the field of Earth remote sensing, to monitor the melting state of surfaces covered by snow or to detect the presence of fresh frost using visible polarimetry (Kelly & Hall, 2008). From Earth orbit, polarimetry is mainly used to retrieve the aerosols properties, but the data inversion requires the correct removal of the contribution of the surface to the polarimetric signal. We note here that inversion of aerosols above land surface covered by fresh frost might have to take into account the large amplitude and narrow angular variations of the degree of polarization induced by the frost.

## Acknowledgments and Data

The construction of the POLICES facility was funded by the National Center for Competence in Research ''PlanetS'' of the Swiss National Science Foundation. We are grateful to all the engineers and technicians of the WP department at the University of Bern who participated in its construction. O.P. thanks Pr. Hans Martin Schmid for fruitful polarimetric discussions. O.P. also thanks the NCCR PlanetS and a post-doctoral fellowship from CNES for funding. We thank Karly Pitman and an anonymous reviewer for their insightful comments that helped to improve this manuscript.

The data files of all the phase curves are available as supporting information (NB: files not available on Arxiv because ancillary files are not supported with PDF submissions, but you can contact the author to obtain the data files).

## Supplementary Materials:

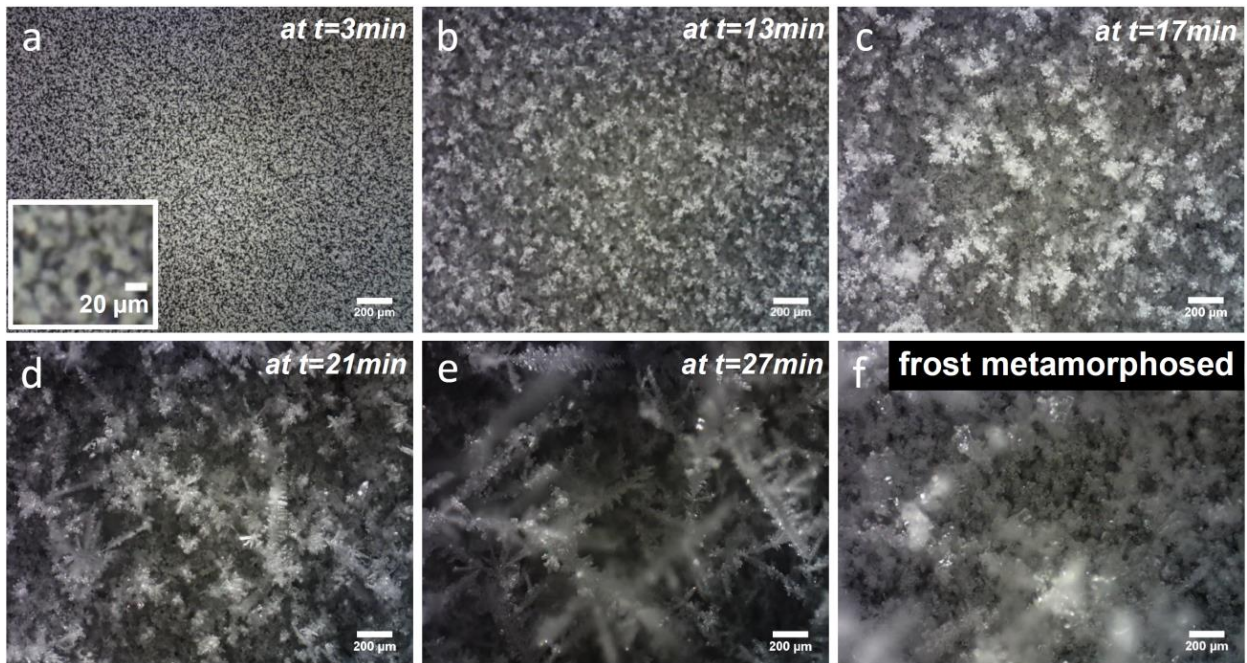

**Figure S1.** Optical microscopy images acquired during the growth and metamorphism of water frost on a dark surface cooled with liquid nitrogen, sample named "frost on dark". **(a)** First, unresolved roundish particles smaller than about 10 µm are formed and agglomerated at the surface. **(b,c)** Then, dendrites start to form on this surface. **(d,e)** With time, dendrite's "trunks" 10 to 20 µm large get 400 µm to 2-3 mm long, while dendrite's branches are 60 to 70 µm long. **(f)** When the frost layer gets thick and its temperature close to 273 K, the particles are transformed in larger crystals more translucent and with angular facets (60, 70 to 150 µm large), while the aggregated particles at the bottom layer get also larger (20 to 30 µm). In order to have most of the frost particles in-focus, extended depth of field images were made using "CombineZP" program developed by Alan Hadley (https://en.wikipedia.org/wiki/CombineZ).

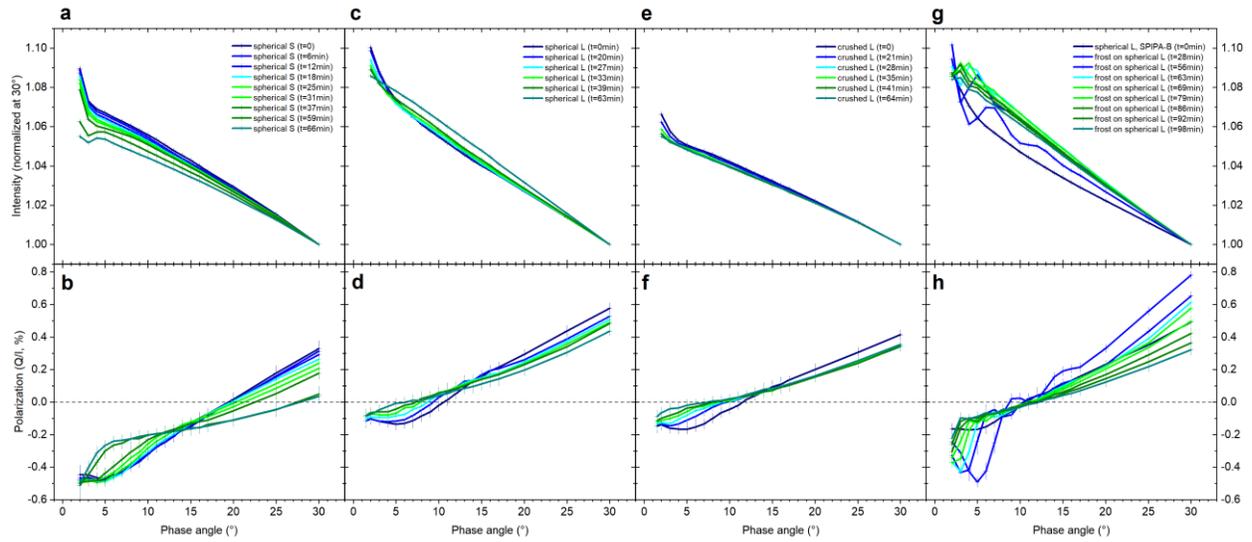

**Figure S2.** Complete temporal evolution and the photometric **(a,c,e,g)** and polarimetric **(b,d,f,h)** phase curves of each sample of ice as its particles sinter with time (and grow for the frost). The progressive change of shape of the negative polarization branch (NPB) together with the vanishing of the surge of brightness at small phase angles are clearly seen. It is also obvious that the samples made of the smallest particles (spherical S and frost) exhibit much lower minima of polarization in the NPB. The polarimetric parameters $P_{min}$, $\alpha_{min}$ and $\alpha_{inv}$ extracted from these data are shown in the parameter plots of Figure 7. All these data are available in supporting information Data Set S1.

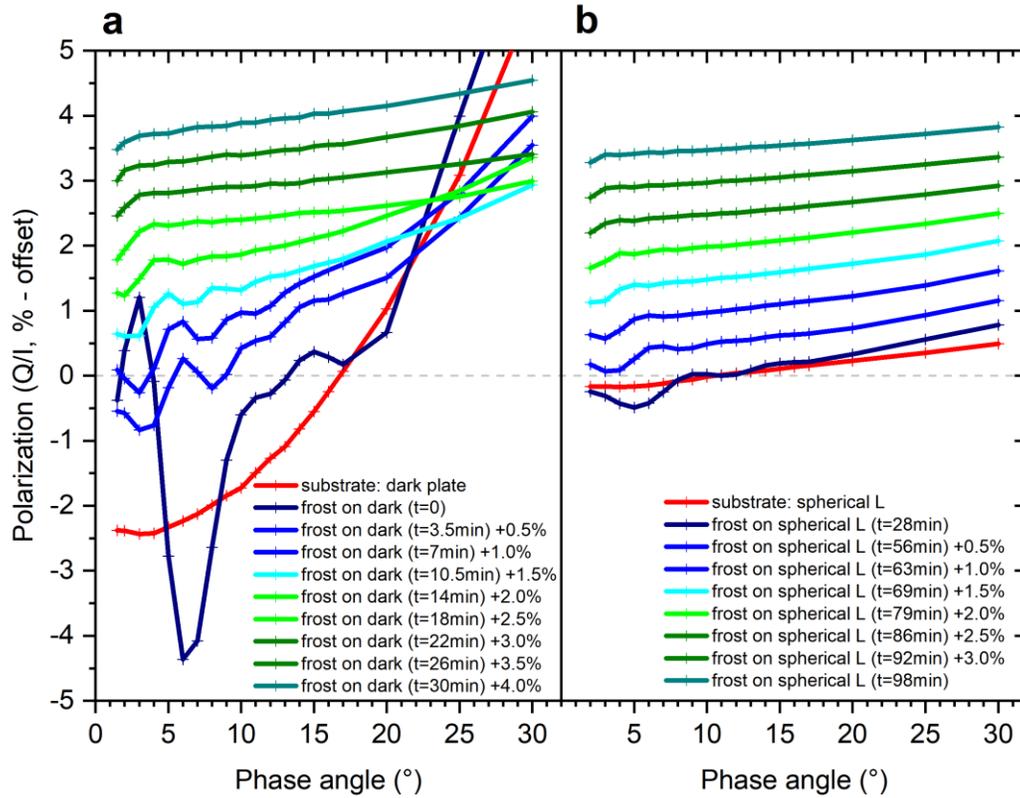

**Figure S3.** Temporal evolution and the polarimetric phase curves of frost growing and sintering on cold surfaces made of **(a)** a dark aluminum tape and **(b)** a surface of spherical L particles. The red curves correspond to the substrates without frost. The polarimetric phase curves of frost are vertically offset (value of the offset indicated in the legend) to allow a better view of the oscillations and their progressive shift toward smaller phase angles. Frost formed on a dark surface exhibits oscillations of much higher amplitude, but similar period and evolution than frost formed on bright spherical L ice grains. The data are available in supporting information Data Set S1.

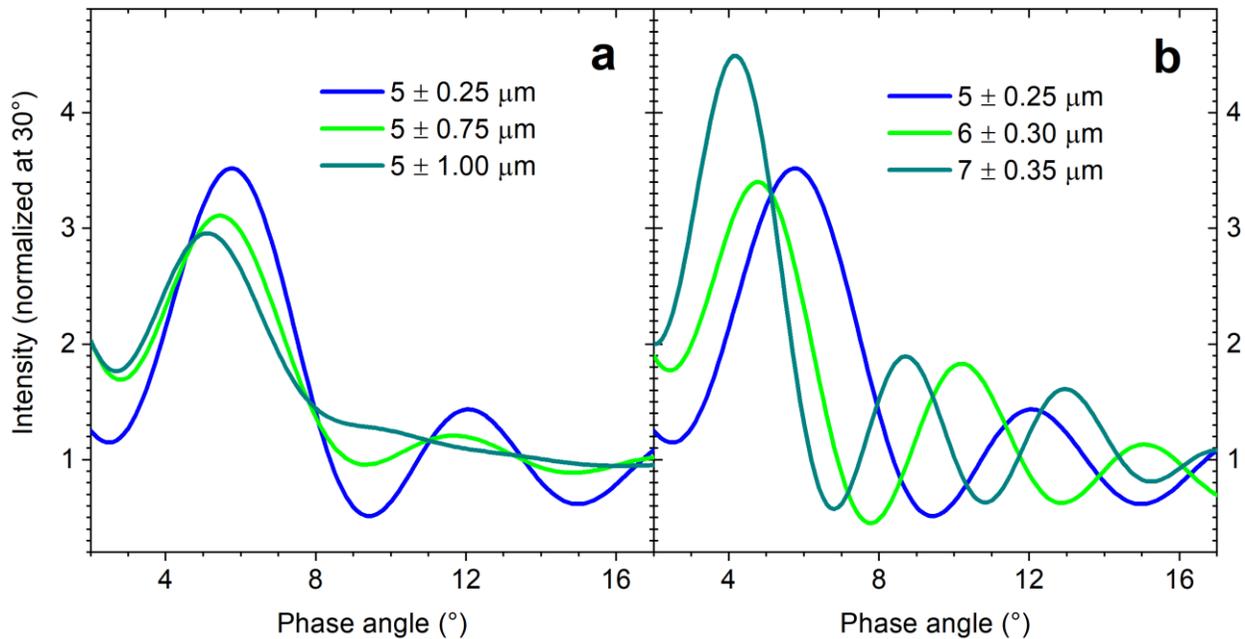

**Figure S4.** Modeled Mie oscillations of brightness computed using the "MiePlot" program developed by Philipp Laven (http://www.philiplaven.com/mieplot.htm). **(a)** As the standard deviation of the size distribution gets larger than 15%, the oscillations decrease in intensity and the secondary maxima vanish. **(b)** As the particle size increases, the oscillations are shifted towards smaller phase angles.

### Data Set S1. (data01.zip)

Data of all the phase curves presented in Figure 2, 4, 5, 6, S2 an S3. The data are presented in files named "phase_curves_[wavelength]nm_[sample name]_[sample time].txt" with [wavelength] = the central wavelength of the incident light (530, 625 or 810 nm), [sample name] = the sample name, as described in Table 1, as well as the photometric standard materials described in section 2.1.2, [sample time] = the time of start of the measurement, counted since the removal of the lid covering the sample (see section 2.1.1). In each file, the first column contains the phase angle (in °), i.e. the angle between the incidence angle (variable) and the emergence angle (fixed at 0°), the following columns contain the intensity (brightness) normalized at 30°, the degree of linear polarization (Q/I, in %) and its standard deviation. Non-numerical values ("ND" for "Not Determined") can be found in the intensity column at low phase angles (1.5°, 2°, 3°, 4° or 5°). This is due to the fact that the goniometer light arm can mask a fraction of the light spot illuminating the sample surface, artificially reducing the light intensity. This is the case at 1.5° and also at 2°, 3°, 4° and 5° when using a depolarizer in front of the incident light beam. However, this "partial eclipse" of the light spot does not prevent to measure the degree of polarization at these phase angles (because it is a ratio of light intensities), so these values are shown in the other columns of the data files.

### Data Set S2. (data02.zip)

Data of all the polarimetric parameters extracted from the phase curves, presented in Figure 7. The data are presented in files named "polarimetric_parameters_[wavelength]nm_[sample name].txt" with [wavelength] = the central wavelength of the incident light (530, 625 or 810 nm), [sample name] = the sample name as described in Table 1. In each file, the first column contains the time of start of the measurement, counted since the removal of the lid covering the sample (see section 2.1.1), and the following columns contain the minimum of linear polarization (Q/I) in the negative branch ($P_{min}$, in %) and its standard deviation, the phase angle at this minimum of polarization ($\alpha_{min}$, in %) and its standard deviation, the phase angle at inversion of the polarization ($\alpha_{inv}$, in %) and its standard deviation.

# References

Blackford, J. R. (2007). Sintering and microstructure of ice: a review. *Journal of Physics D: Applied Physics*, *40*(21), R355–R385. https://doi.org/10.1088/0022-3727/40/21/R02

Breon, F.-M., & Maignan, F. (2017). A BRDF–BPDF database for the analysis of Earth target reflectances. *Earth System Science Data*, *9*(1), 31–45. https://doi.org/10.5194/essd-9-31-2017

Bryant, H. C., & Cox, A. J. (1966). Mie Theory and the Glory*. *Journal of the Optical Society of America*, *56*(11), 1529. https://doi.org/10.1364/JOSA.56.001529

Buratti, B. J., Newman, S., Grundy, W., & Mosher, J. (2008). Correlative Observations of Enceladus and Europa: Clues to Unusual Morphology and Surface Activity. *AGU Spring Meeting Abstracts*, *31*, P31A-04.

Buratti, B. J. (1995). Photometry and surface structure of the icy Galilean satellites. *Journal of Geophysical Research*, *100*(E9), 19061. https://doi.org/10.1029/95JE00146

Cassidy, T. A., Paranicas, C. P., Shirley, J. H., Dalton III, J. B., Teolis, B. D., Johnson, R. E., et al. (2013). Magnetospheric ion sputtering and water ice grain size at Europa. *Planetary and Space Science*, *77*, 64–73. https://doi.org/10.1016/j.pss.2012.07.008

Clark, R. N., Fanale, F. P., & Zent, A. P. (1983). Frost grain size metamorphism: Implications for remote sensing of planetary surfaces. *Icarus*, *56*(2), 233–245. https://doi.org/10.1016/0019-1035(83)90036-2

Clark, R. N., Cruikshank, D. P., Jaumann, R., Brown, R. H., Stephan, K., Dalle Ore, C. M., et al. (2012). The surface composition of Iapetus: Mapping results from Cassini VIMS. *Icarus*, *218*(2), 831–860. https://doi.org/10.1016/j.icarus.2012.01.008

Clarke, D., & Fullerton, S. R. (1996). The Sun as a polarimetric variable star. *Astronomy and Astrophysics*, *310*, 331–340.

Dalton, J. B., Shirley, J. H., & Kamp, L. W. (2012). Europa's icy bright plains and dark linea: Exogenic and endogenic contributions to composition and surface properties. *Journal of Geophysical Research: Planets*, *117*(E03003). https://doi.org/10.1029/2011JE003909

Dollfus, A. (1957). Étude des planètes par la polarisation de leur lumière. *Supplements Aux Annales d'Astrophysique*, *4*, 3–114.

Domingue, D., Hartman, B., & Verbiscer, A. (1997). The Scattering Properties of Natural Terrestrial Snows versus Icy Satellite Surfaces. *Icarus*, *128*(1), 28–48. https://doi.org/10.1006/icar.1997.5733

Dougherty, L. M., & Geake, J. E. (1994). Polarization by frost formed at very low temperatures, as relevant to icy planetary surfaces. *Monthly Notices of the Royal Astronomical Society*, *271*. https://doi.org/10.1093/mnras/271.2.343

Engler, N., Schmid, H. M., Thalmann, C., Boccaletti, A., Bazzon, A., Baruffolo, A., et al. (2017). The HIP 79977 debris disk in polarized light. *Astronomy & Astrophysics*, *607*, A90. https://doi.org/10.1051/0004-6361/201730846

Filacchione, G., Capaccioni, F., Ciarniello, M., Clark, R. N., Cuzzi, J. N., Nicholson, P. D., et al. (2012). Saturn's icy satellites and rings investigated by Cassini–VIMS: III – Radial compositional variability. *Icarus*, *220*(2), 1064–1096. https://doi.org/10.1016/j.icarus.2012.06.040

Gatebe, C. K., Dubovik, O., King, M. D., & Sinyuk, A. (2010). Simultaneous retrieval of aerosol and surface optical properties from combined airborne- and ground-based direct and diffuse radiometric measurements. *Atmos. Chem. Phys.*, *10*, 2777–2794. https://doi.org/10.5194/acp-10-2777-2010

Geake, J. E., & Dollfus, A. (1986). Planetary surface texture and albedo from parameter plots of optical polarization data. *Monthly Notices of the Royal Astronomical Society*, *218*(1), 75–91. https://doi.org/10.1093/mnras/218.1.75

Geake, J. E., & Geake, M. (1990). A remote-sensing method for sub-wavelength grains on planetary surfaces by optical polarimetry. *Monthly Notices of the Royal Astronomical Society*, *245*, 46–55.

Grundy, W. M., Buratti, B. J., Cheng, A. F., Emery, J. P., Lunsford, A., McKinnon, W. B., et al. (2007). New Horizons Mapping of Europa and Ganymede. *Science*, *318*(5848), 234–237. https://doi.org/10.1126/science.1147623

Gundlach, B., Ratte, J., Blum, J., Oesert, J., & Gorb, S. N. (2018). Sintering and sublimation of micrometre-sized water-ice particles: the formation of surface crusts on icy Solar System bodies. *Monthly Notices of the Royal Astronomical Society*, *479*(4), 5272–5287. https://doi.org/10.1093/mnras/sty1839

Hansen, G. B., McCord, T., (2004). Amorphous and crystalline ice on the Galilean satellites: A balance between thermal and radiolytic processes. *Journal of Geophysical Research*, *109*(E1). https://doi.org/10.1029/2003JE002149

Hansen, G., McCord, T., Clark, R., Cruikshank, D., Brown, R., Baines, K., et al. (2005). Ice Grain Size Distribution: Differences Between Jovian and Saturnian Icy Satellites From Galileo and Cassini Measurements. *AGU Fall Meeting Abstracts*, *11*, P11B-0124.

Hansen, G. B. (2009). Calculation of single-scattering albedos: Comparison of Mie results with Hapke approximations. *Icarus*, *203*(2), 672–676. https://doi.org/10.1016/j.icarus.2009.05.025

Helmlinger, M., & Arrecchi, A. (2012). Space Grade Spectralon. *Conference on Characterization and Radiometric Calibration for Remote Sensing (CALCON)*. Retrieved from https://digitalcommons.usu.edu/calcon/CALCON2012/All2012Content/37

Herman, M., Deuzé, J. L., Devaux, C., Goloub, P., Bréon, F. M., & Tanré, D. (1997). Remote sensing of aerosols over land surfaces including polarization measurements and application to POLDER measurements. *Journal of Geophysical Research: Atmospheres*, *102*(D14), 17039–17049. https://doi.org/10.1029/96JD02109

Howett, C. J. A., Spencer, J. R., Pearl, J., & Segura, M. (2010). Thermal inertia and bolometric Bond albedo values for Mimas, Enceladus, Tethys, Dione, Rhea and Iapetus as derived from Cassini/CIRS measurements. *Icarus*, *206*(2), 573–593. https://doi.org/10.1016/j.icarus.2009.07.016

Jaumann, R., Stephan, K., Hansen, G. B., Clark, R. N., Buratti, B. J., Brown, R. H., et al. (2008). Distribution of icy particles across Enceladus' surface as derived from Cassini-VIMS measurements. *Icarus*, *193*(2), 407–419. https://doi.org/10.1016/j.icarus.2007.09.013

Jia, X., Kivelson, M. G., Khurana, K. K., & Kurth, W. S. (2018). Evidence of a plume on Europa from Galileo magnetic and plasma wave signatures. *Nature Astronomy*. https://doi.org/10.1038/s41550-018-0450-z

Jost, B., Pommerol, A., Poch, O., Gundlach, B., Leboeuf, M., Dadras, M., et al. (2016). Experimental characterization of the opposition surge in fine-grained water–ice and high albedo ice analogs. *Icarus*, *264*, 109–131. https://doi.org/10.1016/j.icarus.2015.09.020

Kelly, R., & Hall, D. K. (2008). Remote Sensing of Terrestrial Snow and Ice for Global Change Studies. In *Earth Observation of Global Change* (pp. 189–219). Springer, Dordrecht. https://doi.org/10.1007/978-1-4020-6358-9_9

Kemp, J. C., Henson, G. D., Steiner, C. T., & Powell, E. R. (1987). The optical polarization of the Sun measured at a sensitivity of parts in ten million. *Nature*, *326*(6110), 270. https://doi.org/10.1038/326270a0

Kim, M.-H., Kim, H., Lee, K.-S., & Kim, D. R. (2017). Frosting characteristics on hydrophobic and superhydrophobic surfaces: A review. *Energy Conversion and Management*, *138*, 1–11. https://doi.org/10.1016/j.enconman.2017.01.067

Kolokolova, L., Hough, J., & Levasseur-Regourd, A. C. (2015). *Polarimetry of Stars and Planetary Systems*. Cambridge University Press.

Laven, P. (2005a). Atmospheric glories: simulations and observations. *Applied Optics*, *44*(27), 5667. https://doi.org/10.1364/AO.44.005667

Laven, P. (2005b). How are glories formed? *Applied Optics*, *44*(27), 5675–5683. https://doi.org/10.1364/AO.44.005675

Laven, P. (2012). Rainbows, Coronas and Glories. In W. Hergert & T. Wriedt (Eds.), *The Mie Theory* (Vol. 169, pp. 193–222). Berlin, Heidelberg: Springer Berlin Heidelberg. https://doi.org/10.1007/978-3-642-28738-1_7

Leroux, C., Lenoble, J., Brogniez, G., Hovenier, J. W., & De Haan, J. F. (1999). A model for the bidirectional polarized reflectance of snow. *Journal of Quantitative Spectroscopy and Radiative Transfer*, *61*(3), 273–285.

Li, L., Liu, Z., Li, Y., & Dong, Y. (2017). Frost deposition on a horizontal cryogenic surface in free convection. *International Journal of Heat and Mass Transfer*, *113*, 166–175. https://doi.org/10.1016/j.ijheatmasstransfer.2017.05.058

Ligier, N., Poulet, F., Carter, J., Brunetto, R., & Gourgeot, F. (2016). VLT/SINFONI Observations of Europa: New Insights into the Surface Composition. *The Astronomical Journal*, *151*(6), 163. https://doi.org/10.3847/0004-6256/151/6/163

Lv, Y., & Sun, Z. (2014). The reflectance and negative polarization of light scattered from snow surfaces with different grain size in backward direction. *Journal of Quantitative Spectroscopy and Radiative Transfer*, *133*, 472–481. https://doi.org/10.1016/j.jqsrt.2013.09.010

Lyot, B. (1929). *Recherches sur la polarisation de la lumière des planètes et de quelques substances terrestres*. Paris.

Lyot, B. (1934). Polarisation des petites planètes. *Comptes Rendus de l'Académie Des Sciences*, *199*, 774–782.

Martin, T. Z., Goguen, J. D., Travis, L. D., Tamppari, L. K., Barnard, L., & Doose, L. (2000). Galileo PPR Polarimetric Phase Curves for the Galilean Satellites (Vol. 32, p. 39.04).

Presented at the Bulletin of the American Astronomical Society. Retrieved from http://adsabs.harvard.edu/abs/2000DPS....32.3904M

Mauk, B. H., Hamilton, D. C., Hill, T. W., Hospodarsky, G. B., Johnson, R. E., Paranicas, C., et al. (2009). Fundamental Plasma Processes in Saturn's Magnetosphere. In M. K. Dougherty, L. W. Esposito, & S. M. Krimigis (Eds.), *Saturn from Cassini-Huygens* (pp. 281–331). Dordrecht: Springer Netherlands. https://doi.org/10.1007/978-1-4020-9217-6_11

Mishchenko, M., Rosenbush, V., Kiselev, N., Lupishko, D., Tishkovets, V., Kaydash, V., et al. (2010). Polarimetric remote sensing of solar system objects. *ArXiv Preprint ArXiv:1010.1171*.

Mishchenko, M. I., Dlugach, J. M., Liu, L., Rosenbush, V. K., Kiselev, N. N., & Shkuratov, Y. G. (2009). Direct solutions of the Maxwell equations explain opposition phenomena observed for high-albedo solar system objects. *The Astrophysical Journal*, *705*(2), L118–L122. https://doi.org/10.1088/0004-637X/705/2/L118

Muinonen, K., Mishchenko, M. I., Dlugach, J. M., Zubko, E., Penttilä, A., & Videen, G. (2012). Coherent Backscattering Verified Numerically for a Finite Volume of Spherical Particles. *The Astrophysical Journal*, *760*(2), 118. https://doi.org/10.1088/0004-637X/760/2/118

Muinonen, K., Penttilä, A., & Videen, G. (2015). Multiple scattering of light in particulate planetary media. In L. Kolokolova, J. Hough, & A.-C. Levasseur-Regourd (Eds.), *Polarimetry of Stars and Planetary Systems* (pp. 114–125). Cambridge: Cambridge University Press. https://doi.org/10.1017/CBO9781107358249.020

Nelson, R. M., Boryta, M. D., Hapke, B. W., Manatt, K. S., Shkuratov, Y., Psarev, V., et al. (2018). Laboratory simulations of planetary surfaces: Understanding regolith physical properties from remote photopolarimetric observations. *Icarus*, *302*, 483–498. https://doi.org/10.1016/j.icarus.2017.11.021

Nussenzveig, H. M. (1979). Complex angular momentum theory of the rainbow and the glory. *Journal of the Optical Society of America*, *69*(8), 1068. https://doi.org/10.1364/JOSA.69.001068

Ovcharenko, A. A., Bondarenko, S. Y., Zubko, E. S., Shkuratov, Y. G., Videen, G., Nelson, R. M., & Smythe, W. D. (2006). Particle size effect on the opposition spike and negative polarization. *Journal of Quantitative Spectroscopy and Radiative Transfer*, *101*(3), 394–403. https://doi.org/10.1016/j.jqsrt.2006.02.036

Parretta, A., & Addonizio, M. L. (2015). Optical and Structural Characterization of Diffuse Reflectance Standards. *International Journal of Optics and Applications*, *5*(2), 33–49.

Peltoniemi, J., Hakala, T., Suomalainen, J., & Puttonen, E. (2009). Polarised bidirectional reflectance factor measurements from soil, stones, and snow. *Journal of Quantitative Spectroscopy and Radiative Transfer*, *110*(17), 1940–1953. https://doi.org/10.1016/j.jqsrt.2009.04.008

Peltoniemi, J., Gritsevich, M., & Puttonen, E. (2015). Reflectance and polarization characteristics of various vegetation types. In A. A. Kokhanovsky (Ed.), *Light Scattering Reviews 9* (pp. 257–294). Berlin, Heidelberg: Springer Berlin Heidelberg. https://doi.org/10.1007/978-3-642-37985-7_7

Pitman, K. M., Kolokolova, L., Verbiscer, A. J., Mackowski, D. W., & Joseph, E. C. S. (2017). Coherent backscattering effect in spectra of icy satellites and its modeling using multi-sphere T-matrix (MSTM) code for layers of particles. *Planetary and Space Science*, *149*, 23–31. https://doi.org/10.1016/j.pss.2017.08.005

Piucco, R. O., Hermes, C. J. L., Melo, C., & Barbosa, J. R. (2008). A study of frost nucleation on flat surfaces. *Experimental Thermal and Fluid Science*, *32*(8), 1710–1715. https://doi.org/10.1016/j.expthermflusci.2008.06.004

Poch, O., Pommerol, A., Jost, B., Carrasco, N., Szopa, C., & Thomas, N. (2016). Sublimation of ice–tholins mixtures: A morphological and spectro-photometric study. *Icarus*, *266*, 288–305. https://doi.org/10.1016/j.icarus.2015.11.006

Presley, M. A., & Christensen, P. R. (1997). Thermal conductivity measurements of particulate materials 2. Results. *Journal of Geophysical Research: Planets*, *102*(E3), 6551–6566. https://doi.org/10.1029/96JE03303

Rosenbush, V., Kiselev, N., & Afanasiev, V. (2015). Icy moons of the outer planets. In L. Kolokolova, J. Hough, & A.-C. Levasseur-Regourd (Eds.), *Polarimetry of Stars and Planetary Systems* (pp. 340–359). Cambridge: Cambridge University Press. https://doi.org/10.1017/CBO9781107358249.020

Roth, L., Saur, J., Retherford, K. D., Strobel, D. F., Feldman, P. D., McGrath, M. A., & Nimmo, F. (2014). Transient Water Vapor at Europa's South Pole. *Science*, *343*(6167), 171–174. https://doi.org/10.1126/science.1247051

Schmid, H. M., Bazzon, A., Roelfsema, R., Milli, J., Menard, F., Gisler, D., et al. (2018). SPHERE/ZIMPOL high resolution polarimetric imager. I. System overview, PSF parameters, coronagraphy, and polarimetry. *Astronomy & Astrophysics*. https://doi.org/10.1051/0004-6361/201833620

Scipioni, F., Schenk, P., Tosi, F., D'Aversa, E., Clark, R., Combe, J.-P., & Ore, C. M. D. (2017). Deciphering sub-micron ice particles on Enceladus surface. *Icarus*, *290*, 183–200. https://doi.org/10.1016/j.icarus.2017.02.012

Shkuratov, Y. (2002). The Opposition Effect and Negative Polarization of Structural Analogs for Planetary Regoliths. *Icarus*, *159*(2), 396–416. https://doi.org/10.1006/icar.2002.6923

Shkuratov, Y., & Ovcharenko, A. A. (2002). Polarization of Light Scattered by Surfaces with Complex Microstructure at Phase Angles $0.1°$–$3.5°$. *Solar System Research*, *36*(1), 62–67.

Shkuratov, Y., Bondarenko, S., Ovcharenko, A., Pieters, C., Hiroi, T., Volten, H., et al. (2006). Comparative studies of the reflectance and degree of linear polarization of particulate surfaces and independently scattering particles. *Journal of Quantitative Spectroscopy and Radiative Transfer*, *100*(1), 340–358. https://doi.org/10.1016/j.jqsrt.2005.11.050

Sparks, W. B., Schmidt, B. E., McGrath, M. A., Hand, K. P., Spencer, J. R., Cracraft, M., & Deustua, S. E. (2017). Active Cryovolcanism on Europa? *The Astrophysical Journal*, *839*(2), L18. https://doi.org/10.3847/2041-8213/aa67f8

Spencer, J. R., Tamppari, L. K., Martin, T. Z., & Travis, L. D. (1999). Temperatures on Europa from Galileo Photopolarimeter-Radiometer: Nighttime Thermal Anomalies. *Science*, *284*(5419), 1514–1516. https://doi.org/10.1126/science.284.5419.1514

Steigmann, G. A. (1993). The two faces of Callisto. *The Observatory*, *113*, 70–74.

Stephan, K., Wagner, R., Jaumann, R., Clark, R. N., Cruikshank, D. P., Brown, R. H., et al. (2016). Cassini's geological and compositional view of Tethys. *Icarus*, *274*, 1–22. https://doi.org/10.1016/j.icarus.2016.03.002

Sun, Z., & Zhao, Y. (2011). The effects of grain size on bidirectional polarized reflectance factor measurements of snow. *Journal of Quantitative Spectroscopy and Radiative Transfer*, *112*(14), 2372–2383. https://doi.org/10.1016/j.jqsrt.2011.05.011

Tanikawa, T., Hori, M., Aoki, T., Hachikubo, A., Kuchiki, K., Niwano, M., et al. (2014). In situ measurements of polarization properties of snow surface under the Brewster geometry in Hokkaido, Japan, and northwest Greenland ice sheet. *Journal of Geophysical Research: Atmospheres*, *119*(24), 13,946-13,964. https://doi.org/10.1002/2014JD022325

van de Hulst, H. C. (1947). A Theory of the Anti-Coronae. *Journal of the Optical Society of America*, *37*(1), 16. https://doi.org/10.1364/JOSA.37.000016

van de Hulst, H. C. (1981). *Light Scattering by Small Particles*. Dover Publications.

Verbiscer, Anne J., Peterson, D. E., Skrutskie, M. F., Cushing, M., Helfenstein, P., Nelson, M. J., et al. (2006). Near-infrared spectra of the leading and trailing hemispheres of Enceladus. *Icarus*, *182*(1), 211–223. https://doi.org/10.1016/j.icarus.2005.12.008

Verbiscer, A., French, R., Showalter, M., & Helfenstein, P. (2007). Enceladus: Cosmic Graffiti Artist Caught in the Act. *Science*, *315*(5813), 815–815. https://doi.org/10.1126/science.1134681

Verbiscer, A. J., Helfenstein, P., & Buratti, B. J. (2013). Photometric Properties of Solar System Ices. In M. S. Gudipati & J. Castillo-Rogez (Eds.), *The Science of Solar System Ices* (pp. 47–72). New York, NY: Springer New York. https://doi.org/10.1007/978-1-4614-3076-6_2

Wu, X., Dai, W., Xu, W., & Tang, L. (2007a). Mesoscale investigation of frost formation on a cold surface. *Experimental Thermal and Fluid Science*, *31*(8), 1043–1048. https://doi.org/10.1016/j.expthermflusci.2006.11.002

Wu, X., Dai, W., Shan, X., Wang, W., & Tang, L. (2007b). Visual and Theoretical Analyses of the Early Stage of Frost Formation on Cold Surfaces. *Journal of Enhanced Heat Transfer*, *14*(3). https://doi.org/10.1615/JEnhHeatTransf.v14.i3.70

Yoldi, Z., Pommerol, A., Jost, B., Poch, O., Gouman, J., & Thomas, N. (2015). VIS-NIR reflectance of water ice / regolith analogue mixtures and implications for the detectability of ice mixed within planetary regoliths. *Geophysical Research Letters*, *42*, 6205–6212. https://doi.org/10.1002/2015GL064780

Zaitsev, S. V., Rosenbush, V. K., & Kiselev, N. N. (2013). Database of satellite polarimetry. *Advances in Astronomy and Space Physics*, 109–112.